\documentclass[prb,twocolumn,showpacs,floatfix,superscriptaddress]{revtex4}

\usepackage{graphicx}
\usepackage{amssymb}
\usepackage{amsmath} 
\usepackage{bm}
\newcommand{\A}{\alpha}
\newcommand{\B}{\beta}
\newcommand{\up}{\uparrow}
\newcommand{\dn}{\downarrow}

\newcommand{\scp}[2]{\langle #1 | #2 \rangle}
\newcommand{\trip}[3]{\langle {#1}|{#2}|{#3}\rangle}

\newcommand{\e}[1]{\langle {#1}\rangle}
\newcommand{\be}{\begin{equation}}
\newcommand{\ee}{\end{equation}}

\usepackage{color}
\usepackage{ulem}

\begin{document}

\title{Correlated valence-bond states}
 
\author{Yu-Cheng Lin}
\affiliation{Graduate Institute of Applied Physics, National Chengchi University, Taipei, Taiwan}
\author{Ying Tang}
\affiliation{Department of Physics, Boston University, 590 Commonwealth Avenue, Boston, Massachusetts 02215, USA}
\author{Jie Lou}
\affiliation{Department of Physics, Fudan University, Shanghai, China 200433}
\author{Anders W. Sandvik}
\affiliation{Department of Physics, Boston University, 590 Commonwealth Avenue, Boston, Massachusetts 02215, USA}

\begin{abstract}
We study generalizations of the singlet-sector amplitude-product (AP) states in
the valence-bond basis of $S=1/2$ quantum spin systems.  In the standard AP
states, the weight of a tiling of the system into valence bonds (singlets of two
spins) is a product of amplitudes depending on the length of the bonds. We here
introduce {\it correlated AP} (CAP) states, in which the amplitude product is further
multiplied by factors depending on two bonds connected to a pair of sites (here
nearest neighbors). While the standard AP states can describe a phase
transition between an antiferromagnetic (N\'eel) state and a valence-bond solid
(VBS) in one dimension (which we also study here), in two dimensions it cannot
describe VBS order. With the CAP states, N\'eel--VBS transitions are realized as 
a function of some parameter describing the bond correlations. We here study 
such phase transitions of CAP wave-functions on the square lattice. We find examples
of direct first-order N\'eel--VBS transitions, as well as cases where there is an
extended U($1$) spin liquid phase intervening between the N\'eel and VBS states. In
the latter case the transitions are continuous and we extract critical exponents and
address the issue of a possible emergent U($1$) symmetry in the near-critical VBS.
We also consider variationally optimized CAP states for the standard Heisenberg model 
in one and two dimensions and the $J$-$Q$ model in two dimensions, with the latter 
including four-spin interactions ($Q$) in addition to the Heisenberg exchange ($J$) 
and harboring VBS order for large $Q/J$. The optimized CAP states lead to significantly 
lower variational energies than the simple AP states for these models.
\end{abstract}

\date{\today}


\maketitle

\section{Introduction}
\label {intro}

The valence-bond (VB) basis \cite{pauling33,hulthen38,sutherland88,liang88,beach06,wildeboer11} is ideally suited for describing many different types of ground 
states and low-energy excitations of quantum spin models.\cite{fazekas74,shastry81,anderson87,read89a,bonesteel89,beach09,wang10,banerjee10,banerjee11,tran11,tang11a}
In the case of $S=1/2$ spins in the singlet sector, a basis state corresponds to a tiling of the lattice into bonds connecting pairs of sites forming singlets,
such that each spin belongs to one bond. This basis is overcomplete if bonds of all lengths are included. To describe the ground state of a Hamiltonian with 
bipartite interactions, only bonds connecting sites on different sublattices have to be included---this restricted VB basis exactly reproduces Marshall's sign 
rule \cite{marshall55} for the ground state of such a system. Thus, in this basis the wave function is positive definite and can be sampled using Monte Carlo (MC) 
techniques. We here investigate a class of bipartite correlated VB wave functions which can exhibit valence-bond-solid (VBS) order and related interesting quantum
phase transitions in one and two dimensions. 

In this introductory section we provide some further background and motivation for studying VB states. We review the definition and properties of the well-studied 
Liang-Doucot-Anderson amplitude-product states \cite{liang88} and introduce their more versatile generalizations---the {\it correlated AP} (CAP) states that we 
focus on in this paper. We discuss reasons to study such states in the context of quantum phase transitions from the antiferromagnetically ordered N\'eel state 
into non-magnetic VBS and spin liquid states. 

\subsection{Valence-bond states and Marshall's sign rule}

While some analytical work has been carried out in the VB basis,\cite{shastry81,beach09} in most quantitative calculations MC sampling of the bonds must 
normally be used to reliably evaluate expectation values. Since the basis is overcomplete, the non-negative definiteness 
of the wave function is a requirement to avoid problems due to negative sampling weights (the sign problem). Thus, in most cases VB MC calculations are restricted 
to bipartite (non-frustrated) systems. A two-spin singlet (VB) connecting sites $a$ and $b$ on sublattices A and B is then defined according to the
following phase convention:
\begin{equation}
(a,b)=(\up_a\dn_b-\dn_a\up_b)/\sqrt{2}.
\end{equation}
Marshall's sign rule is then incorporated for any tiling of an even number $N$ of spins into $N/2$ singlets,
\begin{equation}
|V\rangle = |(a_1,b_1)\cdots (a_{N/2},b_{N/2})\rangle,
\end{equation}
i.e., when expressed in the standard basis of $z$ spin components, $|Z\rangle =|S^z_1,\ldots,S^z_N\rangle$,
\begin{equation}
|V\rangle = \frac{1}{2^{N/2}} \sum_{Z} \psi_V(Z)|Z\rangle,
\end{equation}
the sign of a non-zero coefficient $\psi_V(Z)$, i.e., for states with antiparallel spins on each bond, is given by 
\begin{equation}
{\rm sign}[\psi_V(Z)]=\psi_V(Z)=(-1)^{n_{A\dn}},
\end{equation}
where $n_{A\dn}$ is the number of $\dn$ spins on sublattice $A$. The wave function $\psi_0(V)$ of the ground state of such a system 
expressed in the VB basis,
\begin{equation}
|\Psi_0\rangle=\sum_V \psi_0(V)|V\rangle,
\end{equation}
is therefore non-negative. When using MC simulations, e.g., with a variational wave function $|\Psi\rangle$ approximating $|\Psi_0\rangle$,
this is essential, because the overcompleteness implies that the sampling weigh is not $|\psi(V)|^2$, as it would be in a standard orthogonal 
basis, but $\psi(V)\psi(V')\langle V'|V\rangle$ for simultaneously sampled non-orthogonal bra and ket configurations $|V\rangle$ and $\langle V'|$. 
The overlap $\langle V'|V\rangle$ and matrix elements of operators of interest in characterizing the states can be easily 
calculated,\cite{liang88,sutherland88,beach06} as we will discuss later below.

\subsection{Amplitude-product states}

The most commonly used variational states in this context are the AP states introduced by Liang, Doucot, and Anderson.\cite{liang88} 
Here one associates a bond connecting two sites $(a,b)$ with an amplitude $h(a,b)$, which in the case of a translationally-invariant system is only a 
function of the lattice vector ${\bf r}_{ab}$ separating the two sites; $h(a,b)=h({\bf r}_{ab})$. The wave function coefficient for a VB configuration $V$ is then
\begin{equation}
\psi(V)=\prod_{{\bf r}}h({\bf r})^{n_{\bf r}},
\label{psiap}
\end{equation}
where $n_{\bf r}$ is the number of bonds of ``shape'' ${\bf r}$ in the configuration. 

The amplitudes $h({\bf r})$ can be used as variational parameters.
In the original work with the AP states to describe the ground state of the two-dimensional (2D) Heisenberg model,\cite{liang88} only the amplitudes for a small 
number of short bonds were optimized, and different functional forms (exponentially or power-law decaying with the distance $r$) were tested. In later work all 
the amplitudes (on finite lattices) were optimized, leading to relative energy errors (deviations from results from unbiased quantum Monte Carlo, QMC, calculations) of 
less than $0.1\%$.\cite{lou07,sandvik10a} In the optimal state the amplitudes decay asymptotically as $1/r^3$, which is also the result of a
mean-field VB approach.\cite{beach09}

In some cases, if one is just interested in the properties of some class of states without reference to a specific Hamiltonian, the optimization step 
is not needed. This approach has been taken in recent studies of the prototypical resonating VB (RVB) spin-liquid state consisting of the superposition 
of all configurations of the shortest (nearest-neighbor) bonds on the square lattice,\cite{tang11b,albuquerque10,ju12} and also in the presence of some 
fraction of the second bipartite bond (fourth-neighbor).\cite{tang11b} These wave functions, for which the parent hamiltonian was recently identified
(in the case of nearest-neighbor bonds only),\cite{cano10} has exponentially decaying spin correlations but power-law decaying VBS correlations. A phase
transition from the N\'eel state into this kind of spin liquid can be achieved by using amplitudes of the form $h(r) \propto 1/r^\kappa$ and tuning
the exponent $\kappa$ to a critical value.\cite{havilio99,beach09}

\subsection{AP states with bond correlations (CAPs)}

\begin{figure}
\includegraphics[width=8.4cm, clip]{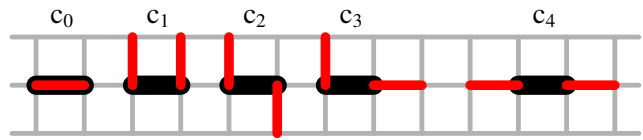}
\caption{(Color online) Configurations of short valence bonds (shown in red) connected to a lattice link $b$ (indicated by thick black bars).
Their associated CAP weights, Eq.~(\ref{psicap}), are $C_b({\bf r}_1,{\bf r}_2)$ with $r_1=r_2=1$ and $b$ here being a horizontal link. We will
later use a notation with weights $c_i$ for CAP states including only short-bond correlations, with $i$ corresponding to $({\bf r}_1,{\bf r}_2)$
according to the labeling above.}
\label{w1234}
\end{figure}

One of the motivations of the work reported in the present paper is to obtain a variational description of the 2D N\'eel--VBS transition. For this, we need a class 
of wave functions beyond the AP states, since they do not exhibit VBS order (while the 1D variants do, as we will discuss in Sec.~\ref{sec:1d}). The 2D non-magnetic 
AP states are beleived to always be spin liquids, with exponentially decaying spin correlations and power-law VBS correlations, similar the prototypical short-bond
RVB states.\cite{beach09,tang11b}

We study a class of generalized AP states defined with bond-correlation factors multiplying the AP wave function (\ref{psiap}). We take these 
factors to be of the form $C_b[{\bf r}_1(b),{\bf r}_2(b)]$, where $b$ denotes a nearest-neighbor link on a 1D chain or 2D square lattice (or, more generally,
any lattice with some imposed bipartition), and ${\bf r}_1(b)$, ${\bf r}_2(b)$ are the shapes of the two VBs connected to this bond (with the case of there 
being just a single bond connecting the two sites being a special case). Thus, the wave-function coefficient is 
\begin{equation}
\psi(V)=\prod_{{\bf r}}h({\bf r})^{n_{\bf r}}\prod_{b}C_b[{\bf r}_1(b),{\bf r}_2(b)].
\label{psicap}
\end{equation}
For a translationally invariant system $C_b({\bf r}_1,{\bf r}_2)$ for given $({\bf r}_1,{\bf r}_2)$ depends only on the orientation (horizontal or vertical in
two dimensions) of the bond $b$, and these weights also should obey applicable lattice symmetries. The number of different correlation factors is then $\propto N^2$ 
for a system of $N$ spins. For simplicity of the notation we hereafter suppress the subscript $b$.

In principle, in variational CAP calculations all correlation factors can be optimized, along with the amplitudes $h({\bf r})$, but one can also opt to consider 
only those factors $C({\bf r}_1,{\bf r}_2)$ for which $r_1,r_2 \le r_{\rm max}$, with some maximum bond length $r_{\rm max}$, and set the remaining weights to unity. 
The possible two-bond configurations with $r_{\rm max}=1$ are illustrated in Fig.~\ref{w1234}. In variational calculations one would expect the energy to decrease 
monotonically with $r_{\rm max}$, which we will demonstrate explicitly in Sec.~\ref{sec:variational}.

Beyond improving the energy in variational calculations, the correlation factors also play an important qualitative role in 2D systems---without bond correlations, the 
standard AP states are either long-range N\'eel ordered (although they do not, by construction, break the spin-rotation symmetry, they can still develop the magnitude of 
the sublattice magnetization) or are RVB spin liquids with critical VBS correlations (as discussed above in the context of the short-bond RVBs). They cannot form VBS 
order. In contrast, the trivial 1D AP state with only short bonds is an extreme case of a two-fold degenerate VBS state with alternating links with or without a VB. 
This kind of long-range order remains stable also in the presence of some fraction of longer bonds, as we will discuss below in Sec.~\ref{sec:1d}. The generalized CAP 
states (\ref{psicap}) can exhibit 2D VBS order if the correlation factors favor such correlations strongly enough. This is true even with correlations only involving 
only the shortest bonds on the square lattice, illustrated in Fig.~\ref{w1234}. The CAP states open the possibility to study the N\'eel--VBS transition in classes of wave 
functions with the bond correlations parametrized in some way, and to carry out improved variational calculations for systems with VBS order or N\'eel states with 
significant VBS fluctuations.

\subsection{Purpose and outline of the paper}

One of our main reasons to study the CAP states here is to investigate their abilities to describe the 2D N\'eel--VBS transition. This transition has received considerable 
interest recently in the context of ``deconfined'' quantum-criticality (DQC).\cite{senthil04a,senthil04b} Following earlier work on VBS states and quantum-critical phenomena 
in antiferromagnets \cite{read89a,chakravarty89,read89b,murthy90,chubukov94} and topological aspects of phase transitions in 3D classical spin systems,\cite{motrunich04} 
Senthil {\it et al.}~proposed \cite{senthil04a,senthil04b} that the 2D N\'eel--VBS transition is of an unusual kind where the standard Landau rule stipulating a generically 
first-order transition between the two ordered states is violated. The two order parameters in the DQC scenario are both consequences of the same underlying more fundamental 
objects---spinons interacting with an emergent U($1$) symmetric gauge field. The spinons condense in the N\'eel state and confine as pairs in the VBS state. Unbiased QMC
studies of $J$-$Q$ models,\cite{sandvik07,melko08,jiang08,lou09,sandvik10,sandvik11} which include certain multi-spin interactions ($Q$) in addition to the standard 
Heisenberg exchange ($J$), are in general in good agreement with the theoretical predictions. Among the most interesting features observed is an emergent U($1$) symmetry 
of the VBS order parameter (presumably reflecting the emergent gauge field---the ``photon'') as the critical point is approached from the VBS side. Moreover, studies of 
SU($N$) generalizations of the $J$-$Q$ model \cite{lou09} and other spin models \cite{kaul12} have allowed direct connections with analytical large-$N$ calculations for the 
non-compact CP$^{N-1}$ field theory argued \cite{senthil04a,senthil04b} to describe the transition.

\subsubsection{Scope of the paper}

We here investigate whether the 2D N\'eel--VBS transition can be correctly captured with a simple ansatz wave-function of the form (\ref{psicap}) with 
fixed single-bond amplitudes (of a power-law form) and continuously varying short-bond correlation weights of the form in Fig.~\ref{w1234}. The result so far
is negative, in the sense that we do not observe the same kind of continuous VBS transition as in the $J$-$Q$ model. Instead, with parameters chosen such that
there is direct N\'eel--VBS transition, we find strong discontinuities. In other cases we find an RVB spin liquid intervening between the ordered phases. Thus,
it still remains an interesting challenge to find a simple VB description of the DQC point. 

Looking at the VBS order-parameter distribution, we do not observe any emergent U($1$) symmetry of the VBS at the continuous VBS to RVB transition,
impliying that this is not a ``deconfined'' transition in this case. Nevertheless, we find interesting scaling
properties of the angular VBS fluctuations, although the length-scale associated with them are not divergent.

In addition to the 2D studies, we also closely examine the N\'eel--VBS transition within the standard AP states in one dimension, with amplitudes $h(r)$ of the form 
$1/r^\alpha$. This transition, which occurs at a critical value of $\alpha$ (which is not universal but depends on the detailed form of the amplitudes for small $r$) 
was previously studied by Beach,\cite{beach09} but only the spin correlations were computed. Here we extract also the VBS correlations and confirm that there is 
a single critical point versus $\alpha$. The exponents are continuously varying, depending on the short-bond amplitudes.

We also report variational calculations including optimization with the CAP states, minimizing the energy for 1D and 2D Heisenberg and $J$-$Q$ models. 
Naturally, bond correlations have a significant improving effect in VBS phases, but they help also to improve the 2D N\'eel state and the critical ground state of
the 1D Heisenberg chain.

\subsubsection{Outline of the paper}

In Sec.~\ref{sec:sampling} we briefly describe the technical aspects of MC calculations with AP and CAP states. In Sec.~\ref{sec:1d} we study the 
N\'eel--VBS transition in 1D AP states, and in Sec.~\ref{sec:2d} we study the more rich set of states and quantum phase transitions in 2D CAP states. In 
Sec.~\ref{sec:variational} we present some 1D and 2D AP and CAP variational calculations (minimizing the energy as a function of the amplitudes and correlation
factors) for prototypical model Hamiltonians with N\'eel and VBS order (the Heisenberg and $J$-$Q$ models), showing how bond correlations improve the
states. We conclude in Sec.~\ref{sec:summary} with a summary and discussion of future prospects.

\section{MC sampling of CAP states and calculation of observables}
\label{sec:sampling}

\begin{figure}
\includegraphics[width=4cm, clip]{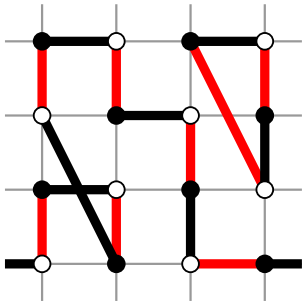}
\caption{(Color online) Transition graph on a $4\times 4$ lattice consisting of two states, $V_{\alpha}$ and $V_{\beta}$, which are depicted by 
red and black bonds, respectively. This transition graph has two loops formed by alternating bonds of $V_{\alpha}$ and $V_{\beta}$.}
\label{fig:transition} 
\end{figure}

In an AP or CAP state with the wave function of the form (\ref{psicap}) the expectation value of an observable $\hat O$ can be written 
for the purpose of importance sampling as
\begin{eqnarray}
\langle \Psi |\hat O|\Psi\rangle & = & 
\frac{\sum_{\A\B} \psi(V_\B) \psi(V_\A) \langle V_\B |\hat O|V_\A \rangle}
{\sum_{\A\B} \psi(V_\B)\psi(V_\A) \langle V_\B |V_\A\rangle} \nonumber \\
&= & \frac{\sum_{\A\B} W_{\A\B}O_{\A\B}}
{\sum_{\A\B} W_{\A\B}},
\label{psiexpvalue}
\end{eqnarray}
where the configuration weight is
\begin{equation}
W_{\A\B} = \psi(V_\B) \psi(V_\A) \langle V_{\beta} |V_{\alpha} \rangle,
\label{weight}
\end{equation} 
and the estimator corresponding to $\hat O$ given by
\begin{eqnarray}
O_{\A\B}=\frac{\langle V_\B |\hat O|V_\A\rangle}{\langle V_\B |V_\A\rangle}.
\end{eqnarray}
Here the overlap $\langle V_\B |V_\A \rangle$ is evaluated by counting the number $l_{\A\B}$ of loops in the transition 
graph of $V_{\A}$ and $V_{\B}$;\cite{sutherland88,liang88}
\begin{eqnarray}
\langle V_\B |V_\A\rangle = 2^{l_{\A\B}}.
\label{overlap}
\end{eqnarray}
An example with two loops is shown in Fig.~\ref{fig:transition}. Below we briefly discuss MC sampling of the VB configurations and
estimators for some important observables.

\subsection{Bond and spin sampling schemes}

For a given functional form for $h({\bf r})$ and bond correlation factors $C({\bf r}_1, {\bf r}_2)$, an expectation value $\e{\hat O}$ can be 
computed stochastically by importance sampling according to the weight $W_{\A\B}$. Using some random reconfiguration of bonds in either the state
$V_\A$ or $V_\B$ or both of them, the standard Metropolis acceptance probability for a modified configuration ($V_{\A'},V_{\B'})$ is
\be
P_{\text{accept}}=\min\left[\frac{W_{\A'\B'}}{W_{\A\B}},1 \right], 
\label{paccept}
\ee 
where the weight ratio is
\begin{equation}
\frac{W_{\alpha^\prime\beta^\prime}}{W_{\alpha\beta}}=\frac{\psi(V_{\A^\prime})\psi(V_{\B^\prime})}{\psi(V_\A)\psi(V_\B)}
2^{(l_{\alpha^\prime\beta^\prime}-l_{\alpha\beta})},
\label{wratio}
\end{equation}
Even for the simplest update involving changes in just two bonds,\cite{liang88} the calculation of the change in the new number of 
loops $(l_{\alpha^\prime\beta^\prime}-l_{\alpha\beta})$ 
can require a computational time up to $\propto N$ for each new update proposal, since a N\'eel state has extensive loops (while magnetically disordered states
have only short loops). Since the number of such updates in each MC sweeps should also be proportional to $N$, this type of update leads to a total computational 
time $\mathcal{O}(N^2)$ for a full sweep in a N\'eel state, while in a non-magnetic state the scaling is $\mathcal{O}(N)$. 

The unfavorable scaling in the N\'eel state can be avoided by working in a combined space of both spins and bonds,\cite{sandvik10a} where the VBs are also sampled, by 
randomly selecting either $\up_a\dn_b$ or $\dn_a\up_b$ for each singlet $(a,b)$. Since the spin basis is orthogonal, all spins in the bra and the ket have to be the same, 
and a consistent assignment for both $V_\alpha$ and $V_\beta$, thus, implies that the spins on each loop in the transition graph follow a staggered, $\up\dn\up\dn\ldots$,
pattern. The overlap (\ref{overlap}) in the pure VB basis then follows, since there are two possible staggered configurations on each loop. The spins are periodically updated 
by flipping all spins in randomly selected loops. By such spin sampling, the weighting by the number of loops is accounted for automatically, due to the entropic effect 
of favoring configurations with large numbers of loops, without the need for actually counting the loops. For a detailed description of the combined spin-bond basis and 
simulations in it, we refer to Refs.~\onlinecite{sandvik10a} and \onlinecite{tang11b}. Here we just note two different ways of updating the bond configurations: 

(i) In the two-bond reconfiguration scheme, an elementary MC move consists of 
choosing two sites on the same sublattice at random (typically the two bonds on a randomly chosen pair of next-nearest-neighbor spins), and exchanging 
the bonds connected to these two sites for the other possible bipartite configuration. Such a reconfiguration is only possible if the spin states on the two 
selected sites are the same. If that is the case, the acceptance probability (\ref{paccept}) is applied, where in the combined spin-bond basis the weight is
\begin{equation}
W_{\A\B} = \psi(V_\B) \psi(V_\A),
\label{weight2}
\end{equation} 
instead of Eq.~(\ref{weight}), and with $W_{\alpha^\prime\beta^\prime}/W_{\alpha\beta}$ evaluated using only the bond amplitudes 
and correlation factors in (\ref{psicap}) affected by the change.

(ii) In a loop update, we start by removing a dimer randomly from two connected sites, creating two defects (``holes''). We keep one defect stationary and move the second one
by connecting one end (the one on the same sublattice) of a chosen bond to it (hence moving the hole to the previous location of that end of the bond). The bond to move 
should be chosen probabilistically in such a way as to satisfy detailed balance, which is relatively straight-forward in the case of AP states \cite{sandvik06,sandvik10a} 
but more complicated when bond correlations are included. In the present work we have used loop updates only for pure AP states, while we use two-bond updates for CAP states.
The latter are also efficient enough to study relatively large lattices (with thousands of spins).

It should be noted here that VB configurations can be classified according to topological ``winding numbers''.\cite{rokhsar88} In AP or CAP states defined with
only short bonds, the two-bond update conserves the winding number, but with no restriction on the bond-length such updates can change the winding number. In
practice, if the bond probability (which depends on the single-bond amplitudes as well as the correlation factors in CAP states) decays very rapidly with the 
length, a simulation for a large system may still be confined to the sector of zero winding number.

\subsection{Spin and dimer correlations}

In order to characterize the different phases realized by the CAP states, we evaluate order parameters for detecting antiferromagnetic (N\'eel) 
order and VBS order. N\'eel order can be characterized using the standard two-spin correlation function,
\be
 C({\bf r}_{ij})=\e{{\bf S}_{{\bf r}_i}\cdot{\bf S}_{{\bf r}_j}}(-1)^{(x_{ij}+y_{ij})},
\label{crr}
\ee
where we use ${\bf r}_{ij}$ to denote the vector separating the lattice sites $i$ and $j$ and the phase factor cancels the signs of the staggered
spin correlations obtaining in the systems we study. Alternatively, one can study the full sublattice magnetization averaged over the whole system;
\be
{\bf m}_s=\frac{1}{N}\sum_{i} \phi_i{\bf S}_{i},
\label{m2def}
\ee
where $\phi_i=+1$ on sublattice A and $\phi_i=-1$ on sublattice B. Since the singlet AP and CAP states manifestly cannot break the spin-rotation symmetry,
order must be detected in the squared order parameter, $\langle {\bf m}_s^2\rangle$, which in the limit of large system size will be identical to the
long-distance spin correlation (\ref{crr}). 

To accurately locate an antiferromagnetic phase transition, the Binder cumulant is very useful. 
It is defined according to \cite{binder81}
\be
     U=\frac{5}{2}\biggl(1-\frac{3}{5}\frac{\e{{\bf m}_s^4}}{\e{{\bf m}_s^2}^2} \biggr),
     \label{bcumulant}
\ee
where the factors are chosen for the $3$-component N\'eel order parameter such that $U(N\to \infty)=0$ in the disordered phase (where the order-parameter distribution
is a Gaussian with zero average) and $U(N\to \infty)=1$ in the ordered phase (where the radial distribution is peaked at non-zero $m_s$). Typically crossing points 
of $U$ graphed versus a control parameter for different system sizes approach the critical point vary rapidly as a function of increasing system size.

To characterize VBS order we use the dimer correlation function, defined as
\be
  \begin{split}
    D_{xx}({\bf r}_{ij}) & =\e{B_x({\bf r}_i) B_x({\bf r}_j)}, \\
    D_{yy}({\bf r}_{ij}) & =\e{B_y({\bf r}_i) B_y({\bf r}_j)},
  \end{split}
  \label{dimercrr}
\ee 
in terms of the bond operators 
\be
   \begin{split}
    B_x({\bf r}_i)& ={\bf S}_{{\bf r}_i}\cdot{\bf S}_{{\bf r}_i+{\hat{\bf x}}},\\
    B_y({\bf r}_i)& ={\bf S}_{{\bf r}_i}\cdot{\bf S}_{{\bf r}_i+{\hat{\bf y}}},
   \end{split}
   \label{bxydef}
\ee
directed along the unit lattice vectors $\hat{\bf x}$ and $\hat{\bf y}$. We will not need the mixed $x$-$y$ 
correlations here. In some cases we will characterize VBS order by the long-distance behavior of (\ref{dimercrr}). The states we will be studying have a 2-site 
VBS unit cell, forming a staggered weak-strong-weak-strong pattern in one dimension and an analogous columnar pattern in two dimensions. In both cases we can 
extract the dominant component of the correlations, corresponding to the squared order parameter, by taking the appropriate difference of (\ref{dimercrr}) 
evaluated at nearby distances. We here use a symmetric version of this difference;
\be
    D^*_{xx}({\bf r})=D_{xx}({\bf r})-\frac{1}{2}\bigl[D_{xx}({\bf r}-\hat{\bf x}) +
                       D_{xx}({\bf r}+\hat{\bf x}) \bigr],
\label{dimercr}
\ee
and a function $D^*_{yy}({\bf r})$ for $y$-oriented dimers defined analogously. We will also study the full order parameter, which in two dimensions can be defined using
the $q=(\pi,0)$ and $q=(0,\pi)$ Fourier transforms of the nearest-neighbor bond correlations (\ref{bxydef});
\be
 \begin{split}
    D_x&=\frac{1}{N}\sum_i (-1)^{x_i} B_x({\bf r}_i),\\
    D_y&=\frac{1}{N}\sum_i (-1)^{y_i} B_y({\bf r}_i).
 \end{split} 
 \label{dxdy}
\ee   
The magnitude $D$ of the order parameter can be computed as the square-root of the average squared operator, 
$\langle D^2\rangle =\langle D_x^2\rangle + \langle D_y^2\rangle$. In addition to the expectation values, we will also investigate the probability 
distribution $P(d_x,d_y)$, in which emergent U($1$) symmetry can be detected. Here $d_x$ and $d_y$ are the expectation values of the corresponding 
operators (\ref{dxdy}) evaluated in a given sampled configuration based on the transition graph. We refer to Ref.~\onlinecite{sandvik12} for further 
details on this quantity, which is not a conventional quantum mechanical expectation value but still very useful for characterising VBS states.

All of the above two- and four-spin correlations are related to the transition-graph loops generated in the VB MC sampling process.
For instance, the estimator for the two-spin correlation is given by \cite{liang88,sutherland88}
\be
\frac{\langle V_\A|{\bf S}_{{\bf r}_i} \cdot {\bf S}_{{\bf r}_j}| V_\B \rangle}{\langle V_\A| V_\B\rangle} = \left \lbrace
\begin{array}{rl}
\pm\frac{3}{4}, & ~~[i,j], \\
0, & ~~[i][j], \\
\end{array}\right.
\label{spincorr}
\ee
where $[i,j]$ and $[i][j]$ denote sites $i$ and $j$ belonging to the same loop and different loops, respectively,
and the sign in the case $[i,j]$ is $+$ and $-$ for spins on the same and different sublattices, respectively.
From Eq.~(\ref{spincorr}) one can also obtain a very simple expression for the estimator for the squared staggered 
magnetization, 
\be
  \frac{\trip{V_\A}{{\bf m}_s^2}{V_\B}}{\scp{V_\A}{V_\B}}=\frac{3}{4}\sum_{\ell=1}^{l_{\A\B}} \mathcal{L}^2_\ell,     
\ee 
where $\mathcal{L}_\ell$ is the size (the number of sites) of loop $\ell$. 

Both the dimer correlation function and the fourth power of the staggered magnetization involve four-spin correlations. Detailed descriptions on how to
calculate these based on the transition graph of two VB configurations can be found in Refs.~\onlinecite{beach06} and ~\onlinecite{tang11b}. Here we only write 
down the expression for the fourth power of the staggered magnetization, needed for the Binder cumulant (\ref{bcumulant}),
\be
  %\begin{split}
   \frac{\trip{V_\A}{{\bf m}_s^4}{V_\B}}{\scp{V_\A}{V_\B}} = 
\sum_\ell \mathcal{L}_\ell^2 + \frac{15}{16}\biggl(\sum_\ell \mathcal{L}_\ell^2 \biggr)^2 -\frac{5}{8} \sum_\ell \mathcal{L}_\ell^4
   %\end{split} 
\ee
which is also solely determined by the sizes of all loops formed in the transition graph. We note that the Binder cumulant of the VBS order parameter is much
more difficult to evaluate, since its definition in analogy with (\ref{bcumulant}) requires eight-spin correlations. While these also in principle 
can be evaluated in terms of the transition-graph loops,\cite{beach06} the expressions are quite complicated to implement in practice and we have not done so.

\section{N\'eel to VBS transition in one dimension}
\label{sec:1d}

In one dimension, the standard AP states given in Eq. (\ref{psiap}) are able to reproduce a N\'eel-VBS transition without correlation factors. 
We will study this 1D transition carefully in this section, using the very efficient loop update of the VB configurations. 

It is natural to study the evolution
of the state as a function of some parameter governing the long-distance behavior of the amplitudes, e.g., using the power law $h(r)=1/r^\kappa$ with
tunable $\kappa$ or an exponential form. Here we will use the power-law. However, it is already known that the nature of the state is not just determined by the 
asymptotic behavior of $h(r)$, but also depends on details of the short-bond weights.\cite{beach09} In addition to the exponent $\kappa$ 
we here tune the shortest-bond amplitude $h(r=1)=\lambda$. The wave function is, thus, explicitly given by
\begin{equation}
\psi(V)=\lambda^{n_{1}(V)}\prod_{r>1} \left (\frac{1}{r^\kappa}\right )^{n_r(V)},
\label{1daps}
\end{equation}
where $n_r(V)$ again refers to the number of bonds of length $r$ in the bond configuration $V$.

\begin{figure}
\includegraphics[width=8cm, clip]{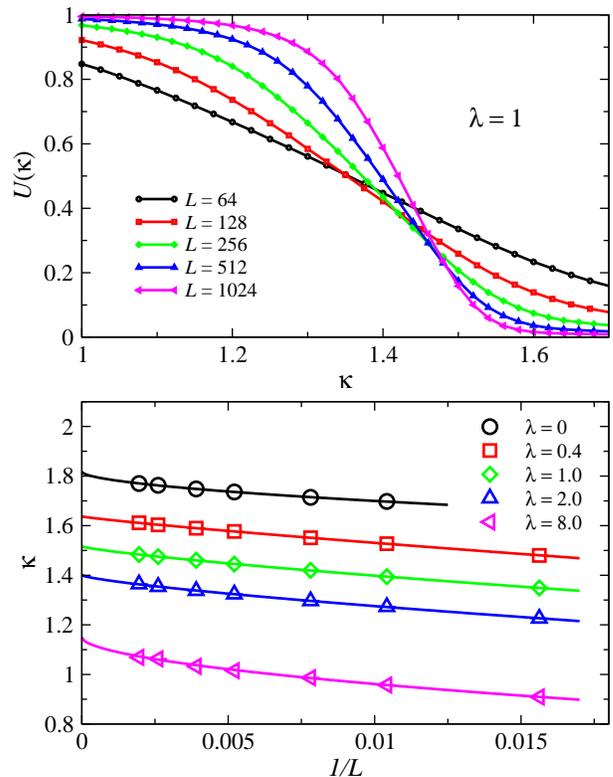}
\caption{(Color online) The upper panel shows the crossing behavior of the Binder cumulant  $U(\kappa)$ defined in Eq. (\ref{bcumulant}) for several
different chain lengths $L$ when $\lambda=1$. The approach to $1$ for small $\kappa$ and $0$ for large $\kappa$ corresponds to the presence and absence of
N\'eel order, respectively. The crossing point approaches the critical value of $\kappa$. The lower panel demonstrates extrapolations to the thermodynamic limit
of the critical $\kappa_c$ by fitting crossing points of ($L,2L$) pairs to the power-law correction (\ref{1dfit}).} 
\label{bindery1}
\end{figure}

It is clear that for $\lambda >0$ and large $\kappa$ this AP state is a VBS, since in the limit $\kappa \to \infty$ only two configurations contribute; those with
$r=1$ bonds on alternating links. For small $\kappa$ there is instead N\'eel order but no VBS order.\cite{beach09} Note that long-range order corresponding to broken 
SU($2$) symmetry is possible in this kind of 1D system, since viewed as a classical statistical-mechanics problem there are long-range interactions (since the bonds
have unbounded length), and the Mermin-Wagner theorem \cite{mermin66} prohibiting 1D N\'eel order does not apply. Note also again that the AP wave function is a singlet 
and, thus, the SU($2$) symmetry is not actually broken (as in any calculation targeting the singlet ground state). The magnitude of the N\'eel order measured by 
$\langle m_s^2\rangle$, Eq.~(\ref{m2def}), or the long-distance correlation function (\ref{crr}) can still evolve toward a non-zero value as the system size grows, 
tending to the square of the symmetry-broken value of $m_s$ in the corresponding thermodynamic-limit state with no constraint on the total spin.

Beach has previously studied N\'eel ordering in this class of wave functions (with a somewhat different parametrization of the short-bond amplitudes).\cite{beach09} 
He found a continuous transition between the N\'eel state and the non-magnetic state. Here we also investigate the VBS correlations and find a single transition 
point where both the spin and dimer correlations are critical. We study the evolution of the transition in the plane ($\kappa,\lambda$). 

\begin{figure}
\includegraphics[width=8cm, clip]{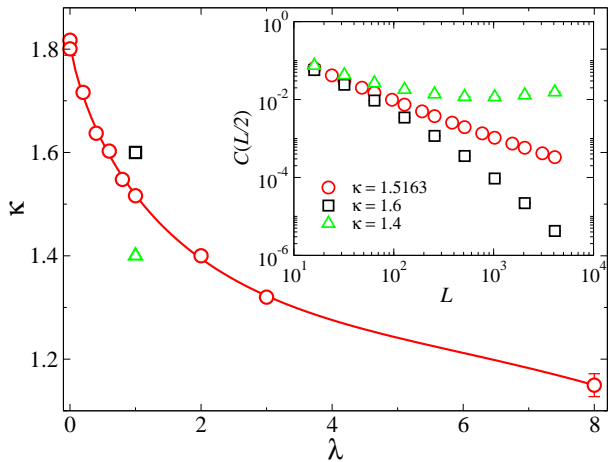}
\caption{(Color online) Phase diagram of 1D AP states with tuning parameters $\kappa$ and $\lambda$, as defined in Eq.~(\ref{1daps}). The circles are calculated 
transition points and the curve is a guide to the eye representing approximately the boundary between the long-range ordered N\'eel (below) and VBS (above) phases. 
The inset exemplifies long-distance spin correlation functions inside the phases and at the critical point when $\lambda=1$; the black squares 
correspond to $\kappa=1.6$ (inside VBS phase) and the green triangles are for $\kappa=1.4$ (in the N\'eel phase). The red circles show the behavior at the 
critical point.} 
\label{1dphase}
\end{figure}

For fixed $\lambda$, in order to find the critical value of $\kappa_c$ of the AP state we study the N\'eel Binder cumulant (\ref{bcumulant}). 
The behavior of curves for different system sizes $L$ crossing each other as a function of $\kappa$ is illustrated in the upper panel of Fig.~\ref{bindery1}. 
The crossing points do not fall exactly on a single point due to subleading size corrections. We observe a systematic smooth drift of the crossing points 
as the system size is increased. In order to eliminate this size effect and determine the critical point from data such as those in Fig.~\ref{bindery1}, 
we extract $\kappa$-values corresponding to crossing points of $(L,2L)$ size pairs, and plot these points against $1/L$, as shown in the lower panel of 
Fig.~\ref{bindery1}. We then extrapolate these values to $L \rightarrow \infty$ and obtain $\kappa_c$. The fitting function we use here for extrapolation 
is the standard power-law;\cite{binder81} 
\begin{equation}
f_c(L, 2L) = \kappa_c + \frac{a}{L^b}.
\label{1dfit}
\end{equation}
The extrapolated $\kappa_c$ values versus $\lambda$ are plotted in Fig.~\ref{1dphase}; the phase diagram of 1D AP states with the two tuning 
parameters $\lambda$ and $\kappa$. The inset of this figure demonstrates the qualitatively different behaviors of the spin correlation functions in the
two phases and at the critical point, using $\lambda=1$ results as an example. At $\kappa=1.6$ the correlations decay faster than power-law, as
is expected for a non-magnetic VBS ordered state. In contrast, at $\kappa=1.4$, the correlations for small $L$ first decay somewhat but then converge 
to a non-zero value for larger $L$ (even increasing somewhat for large systems), demonstrating the presence of long-range N\'eel order. At the critical 
value $\kappa_c$ extracted using the Binder crossings as explained above, the decay of the correlations are consistent with a critical, power-law form.

\begin{figure}
\includegraphics[width=8cm, clip]{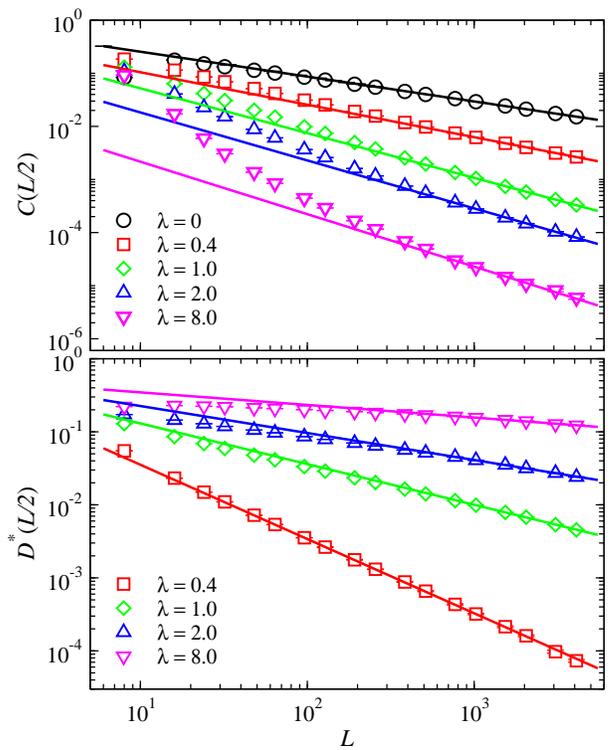}
\caption{(Color online) Staggered spin-spin (upper panel) and dimer-dimer (lower panel) correlations of 1D AP states at the largest distance, 
graphed versus the chain length at $\kappa=\kappa_c$ for different short-bond amplitudes $\lambda$. All lines are fits to the form $aL^{-b}$.}
\label{crry3}
\end{figure}

To determine whether the VBS correlations are also critical at the $\kappa_c$ points extracted from the N\'eel Binder cumulant, we further study both the spin and
dimer correlations at these points. The results confirm the expectation of a common critical N\'eel and VBS point. By studying chains as large as $L=4096$,
we can extract the exponents governing the critical correlation functions with relatively small error bars (thanks to the powerful VB MC loop update discussed 
in Sec.~\ref{sec:sampling}). The analysis of the power laws is presented in Fig.~\ref{crry3}. Note that in order to avoid boundary modifications of the
power-law correlation functions as a function of the distance $r$ in systems of fixed $L$, we study the long-distance correlations versus the system size, with 
$r=L/2$ for the spin correlations and the staggered component of the dimer correlations extracted based on $r=L/2$ and $L/2-1$ data according to 
Eq.~(\ref{dimercr}) [where it should be noted that $D(L/2-1)=D(L/2+1)$ for a periodic chain]. 

As $\lambda$ increases, larger system sizes are needed to observe the asymptotic critical forms. Especially for the largest $\lambda$ studied, $\lambda=8$, one can 
observe in Fig.~\ref{crry3} (upper panel) a clear cross-over of the spin correlation function from a rapidly decaying short-distance form to the asymptotic power-law 
form. The straight lines in Fig.~\ref{crry3} are fits to the simple asymptotic form $aL^{-b}$. We have also tried to include shorter chains in an analysis 
including corrections, by fitting to the form $aL^b+cL^d$. This form is, however, not capable of describing the small size effect in this model (in contrast 
to 2D critical spin liquid RVB states, where this form works very well \cite{tang11b}). In any case, the large-$L$ behaviors appear to be reasonably well 
converged to the simple power law and the exponents extracted should be reliable. An exception is $\lambda=0$, for which the dimer correlations decay very 
rapidly and are too noisy to allow the exponent $\beta$ to be reliably determined (and we have therefore not graphed these correlations in Fig.~\ref{crry3}). 
It is even possible that the VBS state for $\lambda=0$ is of a different kind than for $\lambda>0$. Further studies will be needed to settle this issue.

\begin{figure}
\includegraphics[width=8cm, clip]{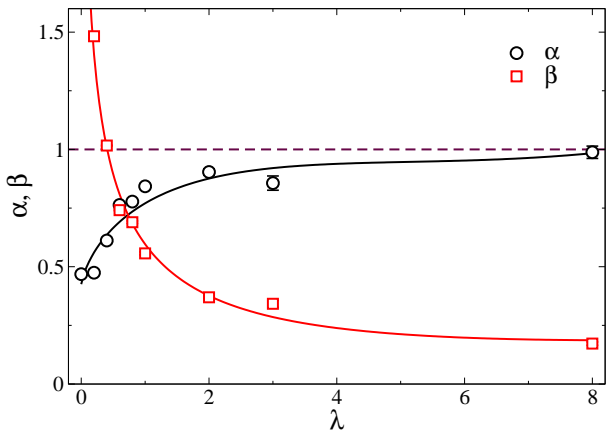}
\caption{(Color online) The continuously varying spin ($\alpha$)  and dimer ($\beta$) decay exponents of the 1D AP state (\ref{1daps}) as a 
function of the short-bond amplitude $\lambda$. The exponents correspond to the power-law decay of the correlation functions; $C(r) \sim r^{-\alpha}$, 
$D^*(r) \sim r^{-\beta}$. The points are calculated values and the curves are guides to the eye.}
\label{exponentsy4}
\end{figure}

We plot the extracted critical exponents as a function of $\lambda$ in Fig.~\ref{exponentsy4}. The exponents vary continuously with $\lambda$, with
the dimer exponent decreasing monotonically and the spin exponent increasing. An interesting conclusion that can be drawn from these results is that 
the critical state becomes increasingly ``quasi VBS ordered'' with increasing $\lambda$, with the decay exponent of the dimer correlations perhaps 
vanishing as $\lambda \to \infty$, although this is difficult to confirm definitely (because the simulations become increasingly difficult for large $\lambda$). 
The behavior is in line with the expectation that a large lambda favors VBS ordering because of the predominance of the very shortest bonds, i.e., when moving
on the critical line toward higher $\lambda$ the density of short bonds increases, and this leads to a strengthening of the VBS quasi-order. At the same time, 
the exponent of the spin correlations appear to approach $1$. However, N\'eel order still exists for large $\lambda$ when reducing $\kappa$ from the critical 
value. In terms of the transition graph estimators of the correlation functions, VBS correlations correspond to certain loop correlations,\cite{beach06} while 
N\'eel order is related to the presence of long ($\propto L$) loops. While long-range N\'eel and VBS orders are mutually exclusive in these states, the N\'eel 
state in the neighborhood of the critical curve for large $\lambda$ approaches a coexistence situation. Here the magnitude of the N\'eel order parameter also 
becomes very small, however, and the coexistence is therefore not robust.

The VB formulation of the ground state can be viewed as a 1D classical statistical mechanics problem, but at the same time it should also correspond to
a path-integral formulation in $1+1$ dimensions (with some underlying parent Hamiltonian). One may then expect the system to be classifiable according to the 
standard 2D conformal field theories by a central charge $c$. Varying critical exponents, as we have found here, normally imply $c\ge 1$, but the fact that the system 
includes long-range interactions may invalidate this requirement, although it is not clear how the power-law bond length translates into effective interactions in an 
underlying parent Hamiltonian (and the interactions in it may well be short-ranged). One notable aspect of the AP states is that they are not able to reproduce the
ground state of the critical Heisenberg chain, where $\alpha=\beta=1$. \cite{affleck85} We will address this issue further in Sec.~\ref{sec:variational} with
variational AP calculations for the Heisenberg chain.

It would be interesting in the future to compute the bipartite entanglement entropy of the 1D AP states, to test its system size scaling and consistency 
between $c$ extracted from it \cite{korepin04,calabrese04} and from the correlation functions. Such calculations can also be carried out using the VB MC 
sampling scheme used here.\cite{ju12,kallin09}

\section{N\'eel to VBS transition in two dimensions}
\label{sec:2d}

The N\'eel state is known to be the ground state of the square lattice Heisenberg antiferromagnet with homogeneous nearest-neighbor couplings.  There 
is convincing numerical evidence\cite{lou07} as well as mean-field arguments \cite{beach09} showing that the standard AP states with power-law decaying 
amplitudes of the asymptotic form $h({\bf r})=1/r^3$ is an optimal variational wave function for the 2D N\'eel state (for any finite size, where the ground 
state is a singlet with no explicitly broke symmetry).

%%%%%%%%%%%%%%%%%%%% FIG %%%%%%%%%%%%%%%%%
\begin{figure}
\includegraphics[width=8cm, clip]{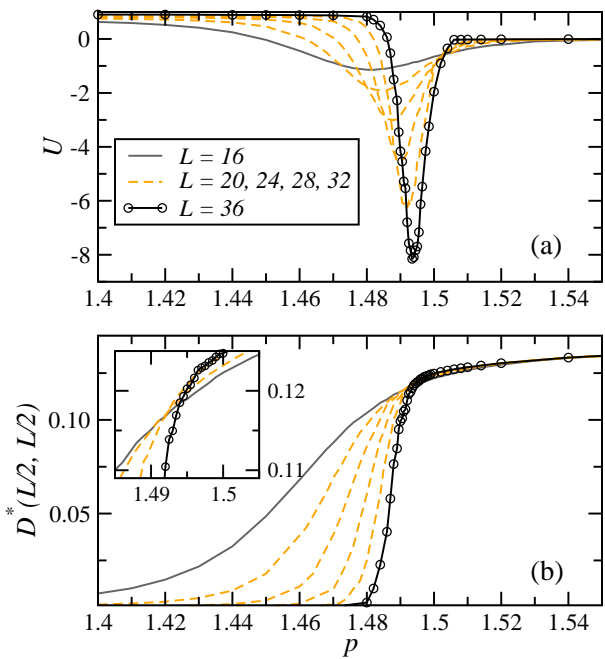}
\caption{\label{fig:1ud}
(Color online) Results for the CAP states with the case (I) parametrization. (a) The Binder cumulant of the staggered magnetization 
as a function of the weight $p$ favoring parallel dimer alignment ($c_1=p$ in Fig.~\ref{w1234}, with the other configurations suppressed 
by setting $c_{i>1}=1/p$). The growing negative peaks of the cumulant indicate the location of the first-order phase transition developing 
as a function of the system size. (b) The columnar component $D^*({\bf r})$ of the dimer correlation function at the largest distance, 
${\bf r}=(L/2, L/2)$, plotted against $p$. The correlation function approaches zero for large systems in the non-VBS state and becomes 
finite in the VBS. The behavior of the correlation function tending to a step-function as $L$ increases, and the curves for different
$L$ crossing each other, is in accord with the first-order phase transition signaled by the pronounced negative cumulant peaks in (a).}
\end{figure}

The AP states have no N\'eel order for rapidly decaying (exponentially or according to a power law $1/r^\kappa$ with large $\kappa$) bonds.\cite{liang88} 
An extreme case is the state that contains only nearest neighbor bonds (dimers). Such a short-bond VB state on the square lattice normally corresponds 
to an U($1$) spin liquid with critical VBS correlations,\cite{tang11b,albuquerque10} in contrast to the 1D AP VBS state discussed in the previous section. 
One should expect the 2D AP state to turn into a long-range ordered VBS when appropriate bond correlations are included. On the square lattice, which we 
will consider here, the simplest kinds of VBS states (with 2-site unit cell) form columnar, staggered, or plaquette ordering patterns. 
  
We here study the N\'eel to VBS transition on a square lattice within the CAP states by imposing bond correlations that favor or suppress only certain types of
short-dimer alignments.  All possible configurations of short dimers connected to a pair of nearest-neighbor sites on a square lattice are shown in Fig.~\ref{w1234}. 
We assign a weight to each of those two-dimer configurations according to Eq.~(\ref{psicap}), with all weights with $r_1,r_2 \not=1$ set to $1$. To simplify
the notation we here use $c_i$ for the short-bond correlation factors, instead of $C({\bf r}_1,{\bf r}_2)$, with the correspondence between the two 
shown in Fig.~\ref{w1234}. For the special case of there being a single bond connecting the two reference sites we set $c_0=1$ as a normalization
factor for the correlations.

To reduce the number of control parameters in our simulations we introduce a single parameter $p$ such that $c_i=p > 1$ for favored two-dimer configurations $i$, 
while $c_i=1/p < 1$ for unfavored configurations and $p=1$ for cases that are considered ``neutral''. If all dimers are uncorrelated, i.e., $p=1$, the state reduces 
to the standard AP state, for which we choose the single-bond amplitudes to be $h({\bf r})=1/r^3$. This choice, which we keep also for the CAP states, is motivated 
by the fact that this gives the correct description of the N\'eel state. The generalized CAP states we use in our simulations are therefore characterized by the
single parameter $p$ controlling the bond correlations.
  
Below we investigate two different parametrizations of the bond correlations. In both cases we use $c_1=p>1$ to locally favor the columnar or plaquette VBS pattern (and 
whichever of these two VBS patterns that actually will be realized is not clear from the outset). Other types of dimer correlations are suppressed, by setting 
$c_2=c_3=c_4=1/p$ in the first case---case (I)---while they are set to neutral, $c_2=c_3=c_4=1$, in case (II). 

We destabilize  N\'eel order by increasing the control parameter $p$ and study the phase transition into a VBS. For case (I), we have found a first-order 
N\'eel to columnar VBS transition, while for case (II) we have found a continuous transition into a critical U($1$) spin liquid, followed by a second continuous 
transition to the columnar VBS. We discuss the two cases in order.

\subsection{A first-order N\'eel to VBS transition} 
\label{sec:first2D}       

%%%%%%%%%%%%%%%%%%%% FIG %%%%%%%%%%%%%%%%%
\begin{figure}
\includegraphics[width=8cm, clip]{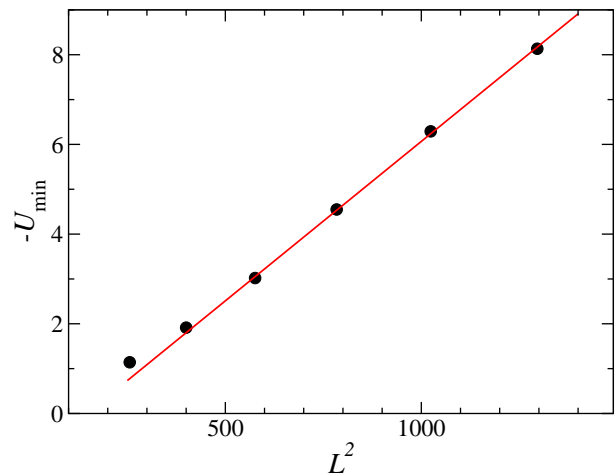}
\caption{\label{fig:1upeak}
(Color online) Finite size scaling behavior of the Binder cumulant minimum of the CAP states, case (I). The negative minimum
increases linearly with $L^2$, indicating a first-order transition in this case.}
\end{figure}

Case (I) again corresponds to favoring parallel VB bond configuration by setting $c_1=p>1$ in Fig.~\ref{w1234} and suppressing fluctuations
by setting $c_{i>1}=1/p$. As in the 1D case discussed in Sec.~\ref{sec:1d}, we here first use the Binder cumulant of the staggered magnetization $U$ to detect the 
transition of the N\'eel order. This quantity is also useful for distinguishing between a first-order and continuous phase transitions. As shown in Fig.~\ref{fig:1ud}(a), 
the Binder cumulant as a function of the control parameter $p$ exhibits a minimum separating the N\'eel phase, where $U\to 1$, and a non-magnetic phase, where $U\to 0$. 
The minimum value of $U$ is negative for all system sizes we studied, and the negative peak becomes narrower and deeper as the system size increases. In fact, the 
negative peak diverges as $-L^2$ when $L\to \infty$, as plotted in Fig.~\ref{fig:1upeak}, which provides strong evidence \cite{vollmayr93} for a first-order 
phase transition. 

Note that the divergence of $U_{\rm min}$ of the form $L^d$ expected for a classical $d$-dimensional system could in principle change to $L^{d+z}$ for a quantum system, where
$z$ is a dynamical exponent.\cite{continentino04} However, our definition of the Binder cumulant is purely a real-space definition and does not include integration over
the imaginary time dimension (which we do not even have access to because it relies on a path-integral formulated using the underlying, unknown parent Hamiltonian). 
The form $L^2$ therefore is expected.

To check whether the non-magnetic phase exhibits VBS order, we next compute the columnar component of the dimer correlation defined according to (\ref{dimercr}). 
Fig.~\ref{fig:1ud}(b) shows $D^*({\bf r})$ for the largest distance, ${\bf r}=(L/2,L/2)$. The correlation indeed converges to a non-zero value in the non-magnetic 
phase, tending to a step function at $p_c$ as $L$ increases. The location of the discontinuity coincides with the point where the Binder cumulant reaches its
minimum in Fig.~\ref{fig:1ud}(a). Note also that the curves for different $L$ cross each other. This size-independence of the order parameter (as supposed to 
size-independence after multiplying with some power of $L$ corresponding to an exponent of critical correlations) at the transition also supports a first-order 
scenario. 

%%%%%%%%%%%%%%%%%%%% FIG %%%%%%%%%%%%%%%%%
\begin{figure}
\includegraphics[width=8cm, clip]{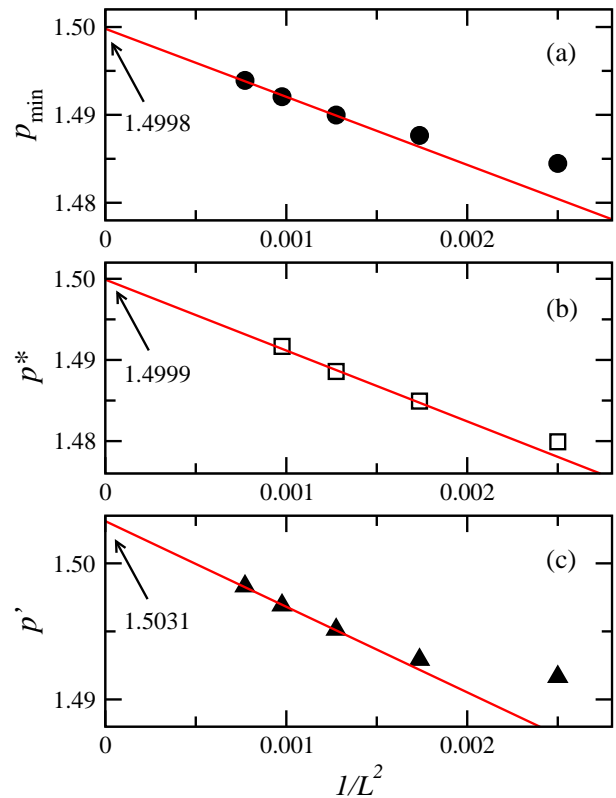}
\caption{\label{fig:1sc}
(Color online) Finite-size scaling plots for extracting the location of the phase transition in CAP case (I).
(a) the location $p_\text{min}$ at which the Binder cumulant reaches its minimum, (b) the crossing point of 
the Binder cumulant for system sizes $(L/2,L$), and (c) the crossing point of the columnar dimer correlation function. 
From finite-size extrapolations of these three quantities, assuming $1/L^2$ dependence, we obtain consistently the 
transition point $p_c$ located within the range  $1.500-1.505$. Note that there are still some visible deviations from
the assumed form and a more precise determination of $p_c$ would require data for still larger systems.}
\end{figure}

The location of the transition point $p_c$ in the thermodynamic limit can be determined, e.g., by extrapolating the $U$ minimum location $p_\text{min}$ (Fig.~\ref{fig:1ud}) 
to the infinite-$L$ limit. The finite-size scaling plot in Fig.~\ref{fig:1sc}(a) shows that the finite-size shift of the transition point defined in this way is consistent 
with $\propto L^{-2}$ for large $L$, where the exponent $2$ again is the one expected based on scaling at a first-order transition, as discussed above. We estimate 
$p_c\approx 1.500$ from an extrapolation to $L\to\infty$. 

In the regime for $p<p_c$ the cumulant for different system sizes exhibits crossing points versus $p$. We expect that the crossing points should coincide with the 
minimum location when $L=\infty$. By finite-size extrapolation of crossing points $p^*$ for pairs of two system sizes $L/2$ and $L$, shown in Fig.~\ref{fig:1sc}(b), 
we estimate $p^*_c\approx 1.500$ in the thermodynamic limit, in perfect agreement with the result obtained from the cumulant minimum. 

In addition to the Binder cumulant signaling the transition of the N\'eel order, we also estimate the transition point of the VBS order from the scaling of the crossing 
points $p'$ of the long-distance dimer correlation function. The inset of Fig.~\ref{fig:1ud}(b) shows a zoom-in of the region where the crossings occur. As shown in 
Fig.~\ref{fig:1sc}(c), these crossing points also appears to shift as $p'\sim L^{-2}$ for large $L$, and the extrapolation to $L\to\infty$ yields an estimated location 
of the transition point $p'_c\approx 1.503$. This is marginally above the two other estimates discussed above, but given the very small range of data points for which 
the $L^{-2}$ fits work well (we have used the points for the three largest systems in all cases, but the data still show some non-asymptotic curvature here), this result 
is still consistent with a single N\'eel--VBS transition point. Larger system sizes would be required to extract the location of this point more precisely.

%%%%%%%%%%%%%%%%%%%% FIG %%%%%%%%%%%%%%%%
\begin{figure}
\includegraphics[width=8cm, clip]{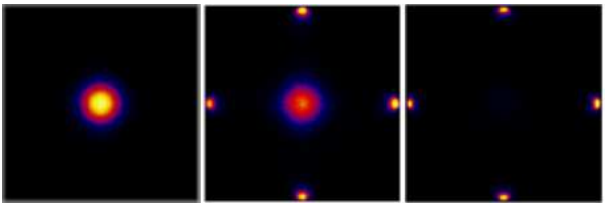}
\caption{\label{fig:1prob}
(Color online) VBS order parameter distribution $P(d_x,d_y)$ for $L=28$ CAP states in case (I).  Brighter regions correspond to higher density.  
Left panel: The distribution at $p=1.46$ is circular-shaped with a central peak, showing U($1$) symmetry in the N\'eel state. Right panel:  
At $p=1.50$ there are four peaks at the Z$_4$-symmetric angles $\phi=\arctan(D_x/D_y)=0,\pi/2,\pi,3\pi/2$, reflecting columnar VBS order. 
Middle panel: At $p=1.48$, in the transition region, the $5$-peak distribution shows coexistence of the N\'eel and VBS order, providing evidence 
for a first-order transition.}
\end{figure}
%%%%%%%%%%%%%%%%%%%%

Finally for case (I) we examine the histogram $P(d_x,d_d)$ of the order parameters $d_x$ and $d_y$, corresponding to VBS order with $x$- and $y$-oriented
dimers, Eq.~(\ref{dxdy}). In Fig.~\ref{fig:1prob} we show $P(d_x,d_y)$ for $L=28$ at $p=1.46, 1.48$ and $1.50$. At $p=1.46$ inside the N\'eel phase, 
the distribution has a circular shape with a central peak. At $p=1.48$ in the transition region, the distribution shows coexistence of the N\'eel order 
(characterized by the central circular region) and the columnar VBS order [characterized by the four narrow peaks at angles $\phi=\arctan(d_y/d_x)=0, 
\pi/2, \pi, 3\pi/2$]; this again provide clear evidence for a first-order N\'eel-VBS phase transition. At $p=1.50$, only barely inside the columnar 
VBS phase of this finite system, the distribution exhibits only the four VBS peaks.

From the many consistent results discussed in this subsection, we can conclude that the CAP states with favored parallel dimer pairs and suppressed flips of such
pairs can characterize a first-order phase transition between the N\'eel and the columnar VBS phases. In such a CAP state, the N\'eel order is destroyed by
formation of parallel dimers as the weight $p$ increases. A first-order transition is also of course what would normally be expected for an order--order transition 
involving two unrelated order parameters. Note also that the system sizes studied in this section were rather modest; up to $L=36$ (while much larger
systems, up to $L=128$, will be considered in the next section). For larger systems it becomes very difficult to obtain good statistics and smooth curves versus 
the control parameter, because of hysteresis effects related to the first-order nature of the transition. Still, as we have shown, the system sizes studied 
are sufficient to study the asymptotic finite-size scalimng forms.

%%%%%%%%%%%%%%%%%%% FIG %%%%%%%%%%%%%%%
\begin{figure}
\includegraphics[width=8cm, clip]{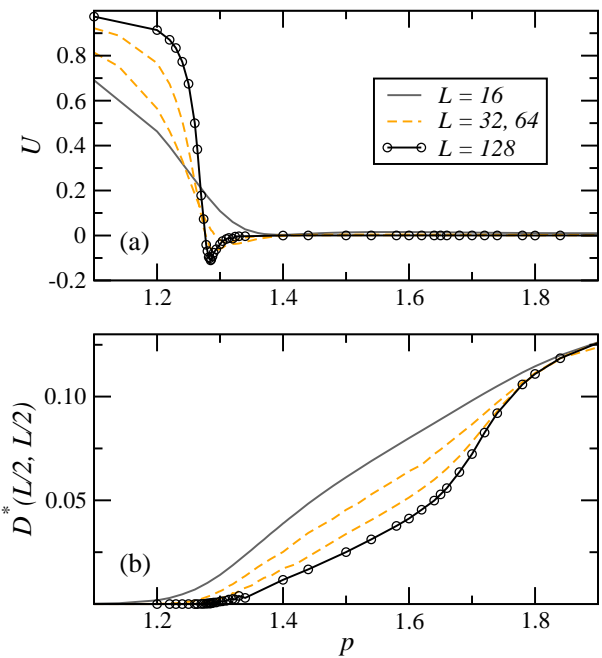}
\caption{\label{fig:2ud}(Color online) Results for CAP states, case (II), for different system sizes plotted versus the control parameter $p$. (a) The 
binder cumulant of the staggered magnetization exhibits a shallow peak at the transition from the N\'eel to a non-magnetic state, with $p_{c1}\approx 1.28$. 
(b) The columnar dimer correlation function indicates a second transition, into the VBS, at $p_{c2} \approx 1.65$, with a completely disordered phase for
$p_{c1} < p < p_{c12}$.}
\end{figure}

In the context of VBS ground states of Hamiltonians, a first-order transition was previously observed with a $J$-$Q$ 
model with the multi-spin interaction $Q$ arranged to favor a staggered state.\cite{sen10} In that case, the first-order transition was expected, because 
local dimer fluctuations are strongly suppressed with this kind of bond order. Alternatively, one can make an argument based on the nature of vortex-like 
defects in the VBS.\cite{banerjee11b} In contrast, in a columnar state parallel short-dimer pairs can fluctuate by $90^\circ$ 
rotation, unless such fluctuations are energetically expensive. In the DQC theory, these fluctuations are essential and correspond to an emergent U($1$) 
symmetry of the VBS order parameter, which has been confirmed in $J$-$Q$ models with plaquette VBS ground states.\cite{sandvik07,jiang08,lou09,sandvik12} 
In the CAP state considered here, we suppressed the fluctuations out of the perfect columnar state by the use of correlation weights and the observed 
first-order transition then is in line with expectations based on the DQC theory and the earlier studies of the N\'eel--VBS transitions in various models.

\subsection{A critical U(1) spin liquid and a second-order transition to the VBS}
\label{sec:second2D}

Given the findings and discussion in the previous section, we now investigate whether the removal of the correlation factors suppressing dimer
fluctuations can change the nature of the N\'eel--VBS transition of the CAP states. In this case (II) we only set $c_1=p>1$ and keep the other correlation 
factors in Fig.~\ref{w1234} as neutral; $c_2=c_3=c_4=1$. 

Fig.~\ref{fig:2ud} shows the N\'eel Binder cumulant and the staggered dimer correlation function against the control parameter $p$. Like in case (I), the 
N\'eel order, characterized by the Binder cumulant tending to $1$ as $L$ grows, survives in the small $p$ region up to a certain value of $p$, and a substantial 
columnar dimer correlation $D^*$ sets in when the N\'eel order is destroyed.  

%%%%%%%%%%%%%%%%%%%% FIG %%%%%%%%%%%%%%%%%
\begin{figure}
\includegraphics[width=8cm, clip]{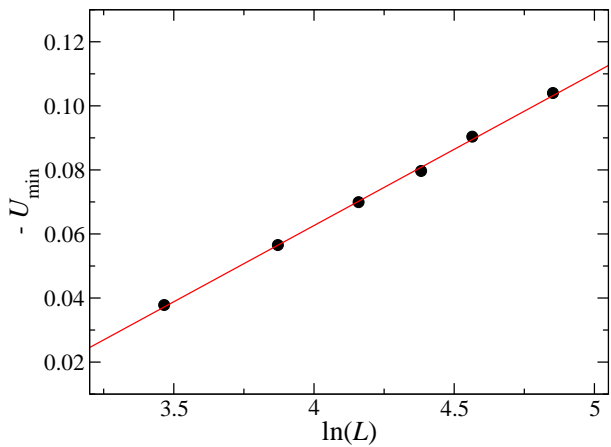}
\caption{\label{fig:2upeak}(Color online)
Finite size scaling behavior of the Binder cumulant minimum in case (II). The negative minimum (obtained by interpolation of the data shown in
Fig.~\ref{fig:2ud}) increases only logarithmically with the system size $L$, indicating a continuous transition.}
\end{figure}

We notice that the negative peak of the Binder cumulant, which occurs only for large systems, is less pronounced than the cumulant peak in case (I).  
A negative Binder cumulant is often taken as evidence for a first-order transition,\cite{cepas05} but there are now known examples of rigorously understood 
continuous classical phase transitions associated with this behavior, e.g., the 2D $4$-state Potts model and the related Ashkin-Teller model in the neighborhood 
of this special point.\cite{jin12} In such cases of ``pseudo-first-order'' scaling the minimum $U_{\rm min}$ of the Binder cumulant diverges much slower 
than the expected $L^d$ form at a first-order transition (or, in some cases, possibly even converges to a finite value) .

The finite-size scaling plot Fig.~\ref{fig:2upeak} shows that $-U_{\rm min}$ for CAP in case (II) grows only logarithmically with the system size.
Thus, we conclude that the N\'eel order here vanishes in a continuous transition associated with pseudo-first-order behavior (related to anomalies
in the critical order-parameter distribution).

%%%%%%%%%%%%%%%%%%%% FIG %%%%%%%%%%%%%%%%%
\begin{figure}
\includegraphics[width=8cm, clip]{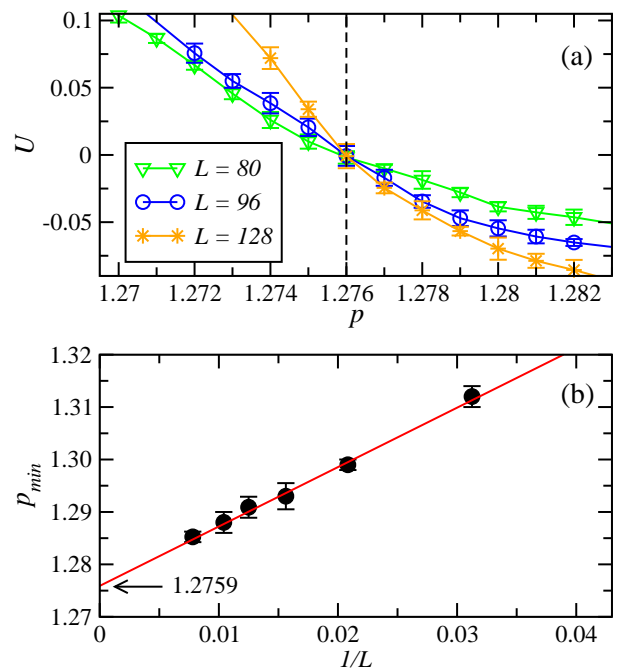}
\caption{\label{fig:2sc}(Color online)
(a) The Binder cumulant for large system sizes in case (II). The curves intersect
at a point, $p \approx 1.276$, which separates the N\'eel phase and the spin liquid.
(b) Finite size scaling of the location $p_\text{min}$ at which the cumulant reaches its 
minimum. The critical point of the N\'eel-spin liquid transition in the thermodynamics 
limit is also here estimated as $p_{1c}\approx 1.276$,  from an extrapolation to $L=\infty$}
\end{figure}

Next we determine the phase boundaries more precisely.  For a continuous phase transition, a frequently used method to determine the critical point is to find 
the unique asymptotic crossing point of the Binder cumulant for different system sizes. Due to finite size effects, the cumulant curves will intersect at a single point 
only when the system sizes are sufficiently large. For our case, we find the intersection point of the cumulant for $L\ge 80$, and it is located at $p\approx 1.276$ 
[Fig.~\ref{fig:2sc}(a)], which can be identified as the N\'eel-spin liquid transition point $p_{1c}$. As the cumulant for a large system size exhibits a negative peak before 
it vanishes in the spin liquid phase, we also extrapolate the location of the peak to $L\to\infty$ to determine the critical point from another route.  By doing so, 
we estimate $p_{1c}\approx 1.276$, in perfect agreement with the cumulant intersection point.    

\begin{figure}
\includegraphics[width=8cm, clip]{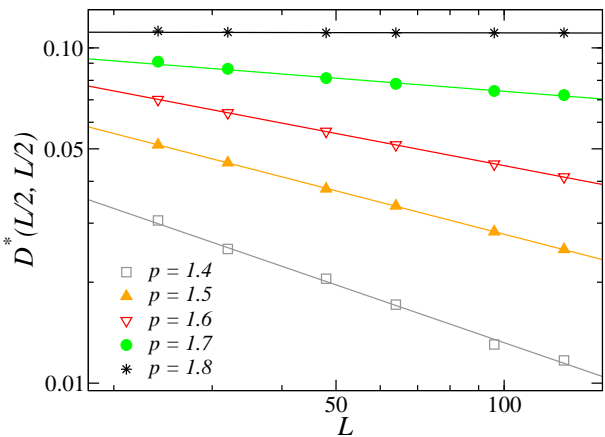}
\caption{\label{fig:ddecay}(Color online)
Long-distance columnar dimer correlation function versus system size for different $p$ in the intermediate region.
the correlations decay as a power-laws (with the decay exponent $\beta= 0.57$ for $p=1.4$, $0.43$ for $p=1.5$,  and $0.32$ for $p=1.6$), in the spin 
liquid phase and are size independent in the VBS state ($p=1.8$). At $p=1.7$ the behavior is approximately a power-law but with a slight up-turn
for the largest size, indicating a weakly VBS ordered state here.}
\end{figure}

Beyond the order of the transition, another major difference from case (I) is the behavior of the dimer correlations. As seen in Fig.~\ref{fig:2ud}(b), there is a wide 
intermediate region between the N\'eel phase and the VBS phase for larger $p$ (where the correlation clearly converges to a non-zero constant for large systems); in this 
intermediate region, the dimer order parameter still decays versus the system size and, for the largest systems, the curve versus $p$ develops a behavior suggestive of 
a second phase transition between a disordered state and the VBS above $p=1.6$. The dimer correlations decay in the intermediate region as a power law, 
$D^*(r) \sim r^{-\beta}$, with the exponent $\beta$ depending on $p$, as shown in Fig.~\ref{fig:ddecay}. Algebraically decaying dimer correlations were 
previously found in short-bond resonating valence bond (RVB) spin liquids investigated in Refs.~\onlinecite{tang11b,albuquerque10}. We thus tentatively identify 
this intermediate $p$-region with critical dimer correlations as a spin liquid in the same class of RVB states, for which it is known that the exponent of
the dimer correlations depends on details of the bond fugacities.\cite{tang11b}  

%%%%%%%%%%%%%%%%%%%%% FIG %%%%%%%%%%%%%%%%%%%%

\begin{figure}
\includegraphics[width=6cm, clip]{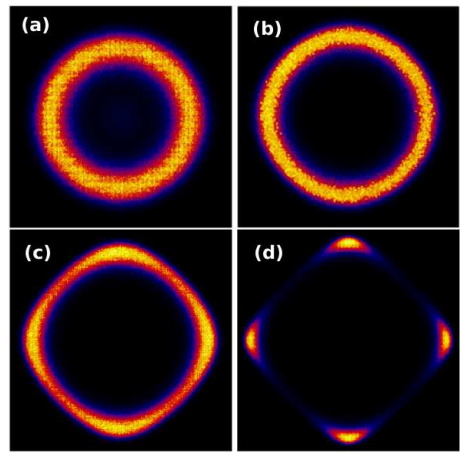}
\caption{\label{fig:2prob}(Color online)
Histograms of the VBS order parameter defined in Eq.~(\ref{dxdy}) shown for
$L=64$ systems in CAP case (II).  In the spin liquid phase with $p=1.4$ and $p=1.5$ [(a) and
(b), respectively], the distributions are ring-shaped with weight at all angles
(bright regions). As $p$ increases to $p=1.6$ (c) and $p=1.7$ (d), the U($1$) symmetric 
distributions evolve into Z$_4$-symmetric ones, with higher densities at the angles $0,\pi/2,\pi,3\pi/2$
corresponding to columnar VBS order.}
\end{figure}

We next study the distribution of the VBS order parameter $P(d_x,d_y)$. The examples of distributions shown in Fig.~\ref{fig:2prob} for $L=64$ are ring shaped
in the intermediate region (at $p=1.4$ and $p=1.5$) before evolving into the expected Z$_4$-symmetric shape in the larger-$p$ regime where columnar VBS order is 
formed (as seen in the location of the peaks; a $45^\circ$ rotated distribution would correspond to a plaquette VBS).  The same kind of ring-shaped $P(d_x,d_y)$ 
distribution was also found in the prototypical short-bond RVB spin liquid (although only when the bond configurations are restricted to the dominant topological 
sector of zero winding number).\cite{tang11b,albuquerque10} 

An important issue here is whether the VBS hosts emergent U($1$) symmetry when the critical point is approached. To investigate this, we also need to
determine the location of the point where the VBS becomes long-range ordered. This is not easy to do just based on the scaling behavior versus the
system size in Fig.~\ref{fig:ddecay}, because there is a whole critical VBS phase and the change from the power-law decay to convergence to a samall non-zero constant
is subtle. We can instead characterize the VBS state by the quantity $\cos(4\phi)$, where $\phi=\arctan(d_y/d_x)$ is the  angle in the VBS order parameter 
($d_x,d_y$) computed for the individual VB configurations, i.e., based on the histogram, $P(d_x,d_y)$. This expectation value measures the degree of the developed 
Z$_4$ symmetry of the VBS order parameter. In the spin liquid (as well as in the N\'eel state), where the U($1)$ symmetry is preserved as $L\to \infty$, we 
have $\e{\cos(4\phi)}=0$. In the VBS states in which the distribution for large systems develops Z$_4$ symmetry, we have $\e{\cos(4\phi)} \to 1$ as $L\to \infty$ 
for a columnar VBS (while it approaches $-1$ for a plaquette VBS). As $\cos(4\phi)$ is dimensionless, one expects $\e{\cos(4\phi)}$-curves for different system sizes 
to asymptotically become size-independent at the transition point into the VBS state. 

%%%%%%%%%%%%%%%%%%%% FIG %%%%%%%%%%%%%%%%%

%%%%%%%%%%%%%%%%%%%% FIG %%%%%%%%%%%%%%%%%
\begin{figure}
\includegraphics[width=8cm, clip]{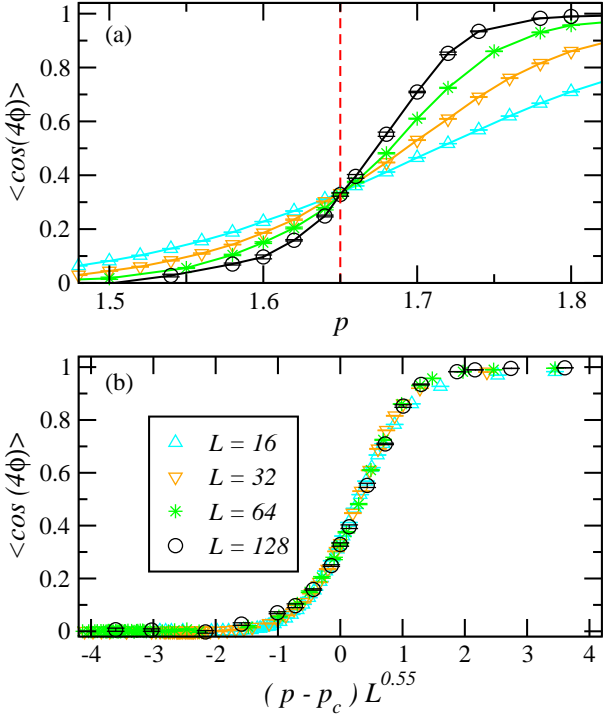}
\caption{\label{fig:4phi}(Color online)
The quantity $\e{\cos(4\phi)}$ with $\phi=\arctan(d_y/d_x)$ measures the degree of $Z_4$ symmetry in the VBS order parameter and also gives a good
estimate of the location of the liquid-VBS transition. In the spin liquid the distribution $P(d_x,d_y)$ is U($1$) symmetric, leading to $\e{\cos(4\phi)}=0$ for 
$L\to\infty$, while in the columnar VBS phase, where the distribution is Z$_4$-symmetric, $\e{\cos(4\phi)}$ approaches $1$. (a) The curves for different system 
sizes cross at a single point located at $p=1.65$, which is identified as the transition point; $p_{2c}=1.650(5)$. Since $\e{\cos(4\phi)} > 0$ at this point, 
there is no emergent U($1$) symmetry at this transition, although it appears to be a continuous transition based on other results. (b) When $p-p_{2c}$ is scaled with
the system size $L$ raised to the power $L^{1/{a\nu}}$, with $a\nu \approx 1.82$, the curves for different systems collapse onto each other.}
\end{figure}

\begin{figure}
\includegraphics[width=8cm, clip]{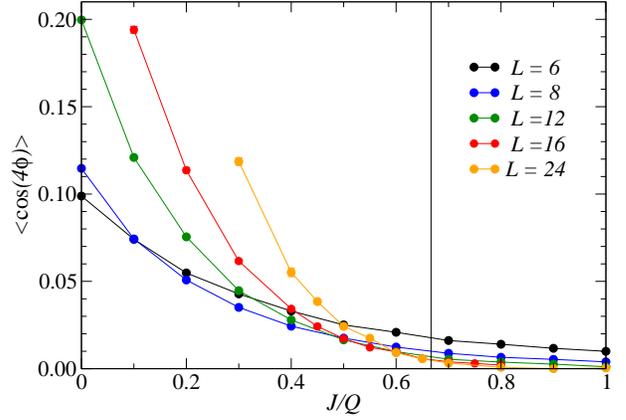}
\caption{\label{jq3cos}(Color online)
The degree of $Z_4$ anisotropy of the VBS order parameter of the $J$-$Q$ model with $6$-spin interactions for different system sizes. Here the curve crossings 
tending to $\e{\cos(4\phi)}=0$ with increasing $L$ demonstrates the emergent U($1$) symmetry at the N\'eel--VBS transition of this model.}
\end{figure}

As seen in Fig.~\ref{fig:4phi}(a), for the case-(II) CAP, a crossing point develops at a non-zero value of $\e{\cos(4\phi)}$, at $p\approx 1.65$, which we thus identify
as the liquid-VBS transition point. The fact that $\cos(4\phi)>0$ at this point shows that there is no emergent U($1$) symmetry at the VBS--liquid transition, since 
the order parameter remains Z$_4$ symmetric exactly at the transition point. For reference, in Fig.~\ref{jq3cos} we show QMC results for the $J$-$Q$ model with 
6-spin columnar $Q$-interactions, for which the order-parameter symmetry was previously analyzed in a slightly different way.\cite{lou09} The results for 
$\e{\cos(4\phi)}$ here show crossing points decaying toward $0$ in the vertical direction. The system sizes are not yet sufficiently large to see that the 
crossings tend toward the critical point, $(J/Q)_c \approx 0.66$. The contrast with the CAP states in Fig.~\ref{fig:4phi}(a) is stark, however, with 
the absence of four-fold symmetry---presence of emergent U($1$) symmetry---at the transition being very plausible.

%%%%%%%%%%%%%%%%%%%% FIG %%%%%%%%%%%%%%%%%

\begin{figure}
\includegraphics[width=8cm, clip]{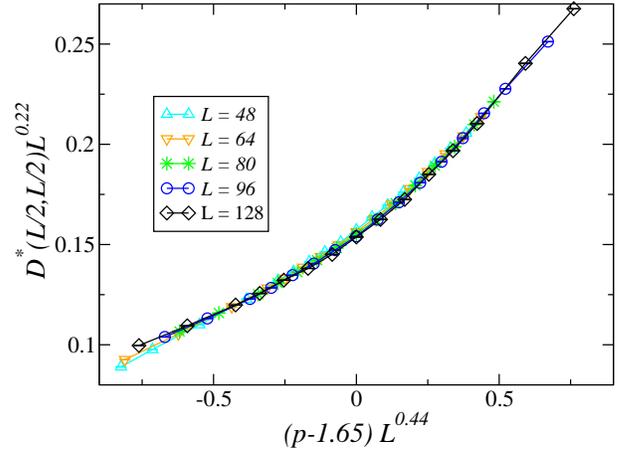}
\caption{\label{dscale}(Color online)
The data of Fig~\ref{fig:2ud}(b) in the neighborhood of the liquid-VBS transition, rescaled to extract the exponent $\beta$ of the dimer correlation function,
$D^*(r) \sim r^{-\beta}$, here with $\beta=0.22$, and the correlation length exponent $\nu$, here with $\nu\approx 1/0.44 \approx 2.27$.}
\end{figure}

Since the results shown in Fig.~\ref{fig:4phi}(a) appear to give a rather precise estimate for the transition point, $p_{2c} = 1.650(5)$, without any scaling 
needed of the vertical axis, we now use this result to investigate scaling of the VBS order parameter. In Fig.~\ref{dscale}, in order to achieve data collapse, 
we have rescaled the long-distance dimer correlation functions in Fig.~\ref{fig:2sc}(b) both vertically, multiplying by $L^\beta$, with $\beta=0.22$, and 
horizontally, multiplying $(p-p_{2c})$ by $L^{1/\nu}$, with $1/\nu=0.44$. The correlation-length exponent is, thus, $\nu \approx 2.3$, which is anomalously 
large and definitely rules out a first-order transition (in which case $\nu=1/d=1/2$ would be expected).

%%%%%%%%%%%%%%%%%%%% FIG %%%%%%%%%%%%%%%%%

The $\e{\cos(4\phi)}$ curves in Fig.~\ref{fig:4phi}(a) can also be scaled in the horizontal direction to collapse all the data onto a single curve, as shown
in panel (b). In cases where there is emergent U($1$) symmetry, this procedure, using the control parameter scaled as $L^{1/{a\nu}}(p-p_c)$, gives the 
correlation-length exponent $\nu$ multiplied by a number $a>1$,\cite{lou09} reflecting the faster divergence of the length-scale $\Lambda$ controlling the 
emergent symmetry; $\Lambda \sim \xi^a$. In the case at hand, we have already concluded that there is no emergent U($1$) symmetry, since $\e{\cos(4\phi)}$
remains non-zero as $p \to p_{2c}$, and the exponent $a\nu$, with $a<1$ should instead reflect a shorter length-scale (irrelevant operator), governing the reduction 
of the angular VBS fluctuations ($\e{\cos(4\phi)} \to 1)$ in the VBS and the growth of these fluctuations ($\e{\cos(4\phi)} \to 0)$ in the liquid. We find very 
good scaling, and indeed the factor $a \approx 0.8$ is clearly less than one.
  
\begin{figure}
\includegraphics[width=8cm, clip]{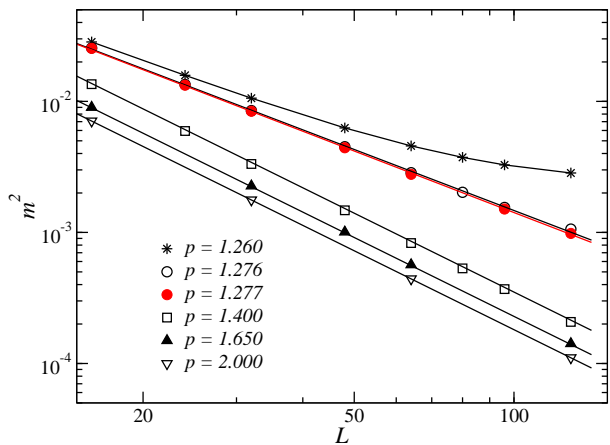}
\caption{\label{fig:m2yc} (Color online)
Finite-size dependence of the squared staggered magnetization for different values of $p$. In the Neel-phase ($p=1.260$), $m^2$ converges to a non-zero value 
for $L\to\infty$. At the Neel-spin liquid  critical point, $p=1.276$, it scales as $m^2 \sim L^{-1.55}$. We see $m^2 \sim L^{-2}$ inside the spin liquid phase 
($p=1.4$), at the liquid-VBS critical point ($p=1.65$), as well as in the VBS phase ($p=2$), indicating an exponentially decaying spin-spin correlation function. 
The lines are fits to the power-law mentioned.}
\end{figure}

We finally investigate the nature of the spin correlations in the spin liquid phase. Results for the squared sublattice magnetization for several points representing
the three different phases are shown in Fig.~\ref{fig:m2yc}. Here we plot the results on a log-log scale, in order to study power-law correlations. In the N\'eel state 
the sublattice magnetization approaches a non-zero constant, while in the liquid and VBS states we observe a clear $1/L^2$ decay. This is the form expected with 
exponentially decaying spin-spin correlation functions. We see a power-law behavior with a non-trivial exponent, $\sim L^{-\alpha}$, with $\alpha=1.55$ 
only at the N\'eel--liquid critical point. These results again confirm that the liquid is of the same type as the prototypical AP RVB states (i.e., with no 
correlation factors), where the varying power-law for the critical VBS correlations corresponds to a varying stiffness constant in a mapping to a 
height model.\cite{tang11b,damle12}
  
\section{Variational calculations}
\label{sec:variational}

In this section we explore variational optimization of CAP states, carrying out energy minimization based on derivatives along the same
lines as in Refs.~\onlinecite{lou07,sandvik10a}. We consider two models. First, the standard Heisenberg model, defined by the Hamiltonian
\begin{equation}
H_J = J\sum_{\langle ij\rangle} {\bf S}_i \cdot {\bf S}_j,
\end{equation}
where $\langle ij\rangle$ denotes a pair of nearest-neighbor sites. We will consider both the 1D chain and the 2D square lattice (in both
cases adopting periodic boundary conditions for systems an even number of spins). We also consider the $J$-$Q$ model,\cite{sandvik07} which includes 
four-spin interactions in addition to the exchange $J$:
\begin{equation}
H_{JQ} = H_J - Q\sum_{\langle ijkl \rangle} ({\bf S}_i \cdot {\bf S}_j-\hbox{$\frac{1}{4}$})({\bf S}_k \cdot {\bf S}_l-\hbox{$\frac{1}{4}$}).
\label{jqham}
\end{equation}
Here $ij$ and $kl$ form opposite edges on an elementary $2\times 2$ plaquette on the square lattice and the summation includes both horizontal
and vertical orientations of these edges on all plaquettes (i.e., the Hamiltonian obeys all the symmetries of the square lattice). With the
negative prefactor of the $Q$ term, this interaction clearly is related to enhancement of the parallel-dimer weight $c_1$, in the notation of Fig.~\ref{w1234}, 
in the ground state wave function (although the state is still of course not expected to be exactly reproduced by the CAP ansatz).

%%%%%%%%%%%%%%%%%%%% FIG %%%%%%%%%%%%%%%%%

\subsection{Optimization method}
\label{sec:optmethod}

We start a variational calculation from some initial value of the parameters in the CAP state (\ref{psicap}), typically a power-law form
of the amplitudes $h$ and all the correlation constants $C=1$. When optimizing states for different values of some parameter (e.g., $J/Q$), 
we also normally start the calculation for a new parameter value from a previous calculation for some nearby value. One can also use this
approach for different system sizes, although when increasing the system size initial values for the parameters corresponding to the 
longest bonds are of course not available and have to be set to some suitable values based on the longest previous bonds. In general, we 
have found that the starting point does not play an important role in optimization of AP and CAP states, indicating that the
energy landscape is relatively smooth.

To minimize the energy as a function of all parameters, we compute the energy and its derivatives. We then apply either (a) the steeped-decent
method or (b) a stochastic variant of it where only the signs of the derivatives are used, as discussed in Ref.~\onlinecite{lou07}. A generic parameter 
$p$ is in these two cases updated according to
\begin{eqnarray}
(a)~~~p & \to & p -\delta \cdot R \cdot {\rm sign}(dE/dp), \nonumber \\
(b)~~~p & \to & p -\delta \cdot (dE/dp)/{\rm max}(|dE|). \label{deltamove}
\end{eqnarray}
Here, in (a) $R \in [0,1)$ is a random number and in (b) ${\rm max}(|dE|)$ denotes the derivative that is the largest in magnitude among all the derivatives considered. 
The maximum shift $\delta$ is gradually reduced, so that the variational parameters eventually will converge. If $\delta$ is reduced sufficiently slowly, then one 
will reach a minimum. This minimum is not necessarily the global one, however. The stochastic scheme (a) should be better in avoiding local minimums, although 
one can of course never be completely guaranteed to find the global minimum. For the case at hand, the energy landscape appears to be relatively smooth, 
with no serious problems in consistently reaching the same minimum energy (regardless of the starting point, as mentioned above). Occasionally the method 
does fail, with independent runs of the same system leading to different final results. Typically, in a set of several runs, most of them will be consistent 
with each other, with only a small fraction of them deviating significantly from the majority value. As expected, the failure rate decreases when increasing the 
number of MC sweeps for sampling the VB configurations (leading to smaller error bars on the derivatives) and when reducing $\delta$ at a slower rate. 

\begin{figure}
\includegraphics[width=7.5cm, clip]{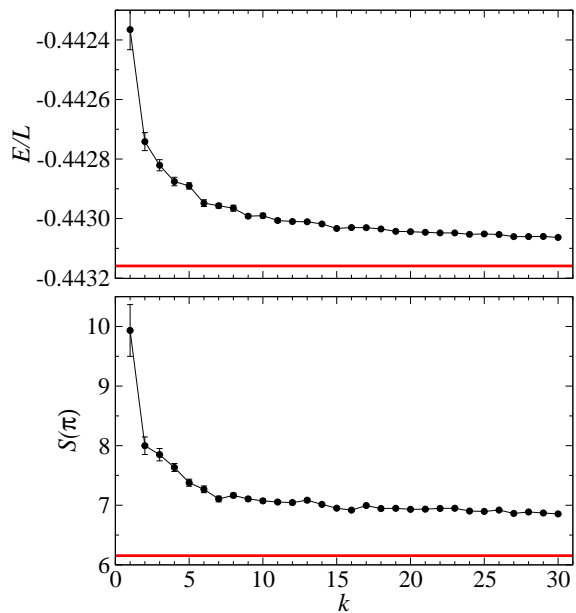}
\caption{\label{conv1d}(Color online)
Convergence of the energy per spin (top panel) and the staggered spin structure factor (bottom panel) as a function of the iteration number $k$ 
for a 1D Heisenberg chain with $L=256$ sites and all correlation amplitudes (i.e., $r_{\rm max} = L/2-1$) included in the CAP wave function (\ref{psicap}). 
For each $k$, $100$ adjustments of the parameters were carried out, each based on an MC simulation with $10^4k^2$ updating sweeps. The points with
error bars correspond to averages over the last $20$ iterations for each $k$. The horizontal lines are results of unbiased QMC calculations.\cite{sandvik10b}}
\end{figure}

We typically use a protocol based on an iteration number $k$. For each $k$, the parameters are adjusted some number $M$ of times (e.g., $M=100$) based on derivatives 
obtained in MC simulations with $\propto k^2$ steps. This way, the derivatives become more precisely determined as the solution approaches the minimum (where
the derivatives decrease in magnitude, thus necessitating a larger number of MC sampling steps to obtain statistically useful information). The maximum 
parameter shift $\delta$ in Eqs.~(\ref{deltamove}) is of the form $\delta_0/k^\alpha$, with $\alpha=3/4$ a suitable exponent in practice, based on experience.

Fig.~\ref{conv1d} shows an example of the evolution of the energy and the sublattice magnetization squared in a run for a Heisenberg chain of length $L=256$.
It is clear that at the final step shown here, $k=30$, the calculation has not yet completely converged (but note that one should not expect convergence to the
shown exact values, as the CAP states still are of course not sufficiently flexible to reproduce the true ground state  wave function completely). In practice, 
one can quite easily converge calculations for small systems essentially completely, while large systems require very long runs. There is still some room for 
improvement of the energy minimization protocol, as what we have explored so far are essentially schemes based on trial-and-error approaches. Nevertheless, the 
results to be presented next can be considered as almost optimized and we do not anticipate that our conclusions would change based on more complete optimizations. 

\subsection{Heisenberg chain}

An interesting question in one dimension is whether AP or CAP states can describe the critical ground state of the Heisenberg chain. As we saw in Sec.~\ref{sec:1d}, 
with the simple parametrized 1D AP wave function (\ref{1daps}) the correct critical decay exponents ($\alpha=\beta=1$) \cite{affleck85} corresponding to this system cannot 
be achieved. In a variationally optimized state, we are not tied to any particular form of the amplitudes, and a sufficiently flexible variational wave function should then 
be able to capture the correct criticality, including the logarithmic corrections that arise in the field-theory language due to a marginally irrelevant operator. The question
then is whether the AP or CAP states have this kind of flexibility, to possibly capture even such a subtle effect as the logarithmic corrections to the correlation
functions.\cite{logpapers} 

To answer the above question, we have carried out energy minimizations with the simple AP state (with all amplitudes as adjustable parameters) as well as with
two types of CAP states. To include only the minimum amount of bond correlations beyond the AP state we include in (\ref{psicap}) only the two 1D bond-pair configurations 
with length-1 bonds; $C(r_1,r_2)$ with $r_1=\pm 1$ and $r_2=\pm 1$. The case $r_1=1,r_2=-1$ corresponds to a single bond on a nearest-neighbor link and we can regard this 
as a normalization for the correlation factors, $C(1,-1)=1$, in the same way as we also set $h(r=1)=1$. There is then only one other correlation factor, $C(-1,1)$ to optimize 
at this level. We also consider the extreme case of optimizing all $C(r_1,r_2)$, $r_{\rm max}=L/2-1$, again with $C(1,-1)=1$.

Let us first discuss the energy. As an example, for $L=256$ the exact energy per site is $E/L=-0.44316$ while for the AP state we obtained $-0.44184$. For the CAP the
best energy when $r_{\rm max=1}$ is $E/L=-0.44272$, and with $r_{\rm max}=L/2-1$ it decreases to $-0.44306$. As discussed in the previous section, it is difficult to
completely optimize long chains, so the optimal variational energies may still be somewhat lower. Following the trends as a function of system size, the relative
energy error with the CAP state seems to remain at about $0.05\%$.

\begin{figure}
\includegraphics[width=8cm, clip]{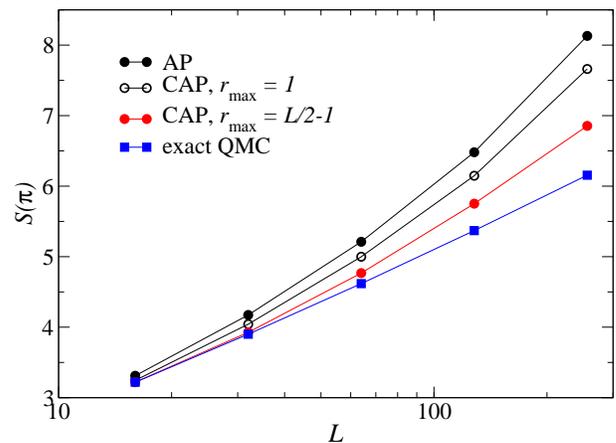}
\caption{(Color online) Staggered spin structure factor of the Heisenberg chain versus system size in variational AP and CAP calculations at 
different correlation levels. The results are compared with exact results from unbiased QMC calculations.\cite{sandvik10b}}
\label{spi}
\end{figure}

Turning now to the spin correlations, in Fig.~\ref{spi} the staggered spin structure factor, defined according to
\begin{equation}
S(\pi)=\sum_{r=0}^{L-1} (-1)^r C(r),
\end{equation}
is graphed versus the system size for all the cases discussed above. Since the exact $C(r)$ decays as $1/r$ with a multiplicative logarithmic correction,
the exact $S(\pi)$ grows slightly faster with $L$ than $\ln(L)$, as demonstrated with unbiased QMC results in Fig.~\ref{spi}. All the AP and CAP results
exhibit a faster growth with $L$. When graphed on a log-log scale (instead of the lin-log scale used in Fig.~\ref{spi}), the AP behavior is consistent
with a power law, $S(\pi) \sim L^{1-\alpha}$ with $\alpha\approx 0.70$. With the CAP states, the data move closer to the exact points, but even with the
maximally correlated CAP state the divergence is still somewhat too fast.

\begin{figure}
\includegraphics[width=8cm, clip]{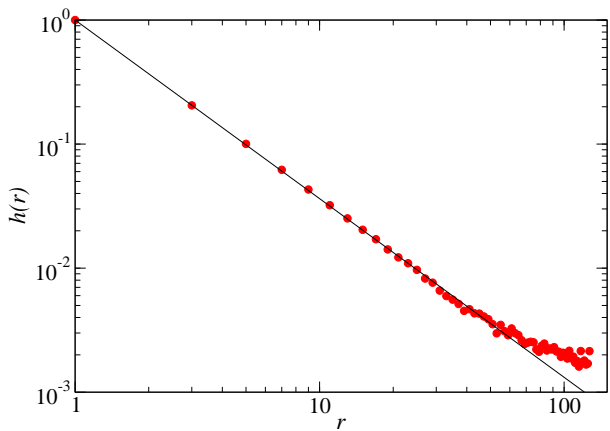}
\caption{(Color online) Optimized AP amplitudes for the Heisenberg chain of length $L=256$. The line has slope $-1.44$.}
\label{hr}
\end{figure}

It is also interesting to examine the optimized amplitudes of the AP state. Fig.~\ref{hr} shows results for $L=256$. Interestingly, a power law applies
here for short and moderate bond lengths, with the deviations (enhancements) at large lengths likely related to the periodic boundary conditions (and some
jaggedness of the large-$r$ data due to imperfect optimization, reflecting the total energy not being very sensitive to these ``noise'' features). Even the $r=1$ 
amplitude falls on the common power-law line in Fig.~\ref{hr}, i.e., in the notation of Sec.~\ref{sec:1d} the optimized state has $\lambda=1$. Looking
at Fig.~\ref{exponentsy4}, when $\lambda=1$ the exponent $\alpha \approx 0.75$, quite close to $\alpha \approx 0.70$ obtained above with the optimized amplitudes. 
Thus, the boundary effects on $h(r)$ seen in Fig.~\ref{hr} appear to have only minor effects on the critical behavior. The conclusion for the optimized AP
state is, thus, that a critical behavior is reproduced, but with the wrong exponents for the correlation functions. Note, however, that $\alpha \approx \beta$
for the applicable power-law obtained here, which is also the case for the true Heisenberg correlations (but with larger values, $\alpha=\beta=1$).

\subsection{Two dimensions}

\begin{figure}
\includegraphics[width=5.5cm, clip]{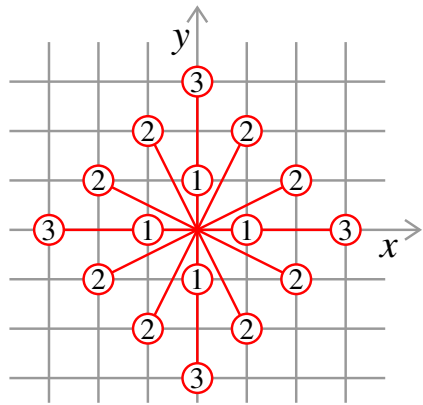}
\caption{\label{levels}
(Color online)  Levels of bond correlations. At level $n$, the longest bonds $({\bf r}_1,{\bf r}_2)$ for which the correlation weight $C({\bf r}_1,{\bf r}_2)$ 
in Eq.~(\ref{psicap}) is optimized (i.e., can be different from $1$) are those marked by $n$.}
\end{figure}

We next systematically investigate the improvement of the energy with the inclusion of bond correlations in two dimensions, using several choices for the 
maximum bond-length $r_{\rm max}$ in the correlation factors $C({\bf r}_1,{\bf r}_2)$. Fig.~\ref{levels} illustrates all the bond shapes $({\bf r}_1,{\bf r}_2)$ 
at three {\it correlation levels}, with $r_{\rm max}=1,\sqrt{5}$, and $3$ for correlation levels 1,2, and 3, respectively.

\subsubsection{Heisenberg model}

\begin{figure}
\includegraphics[width=8.4cm, clip]{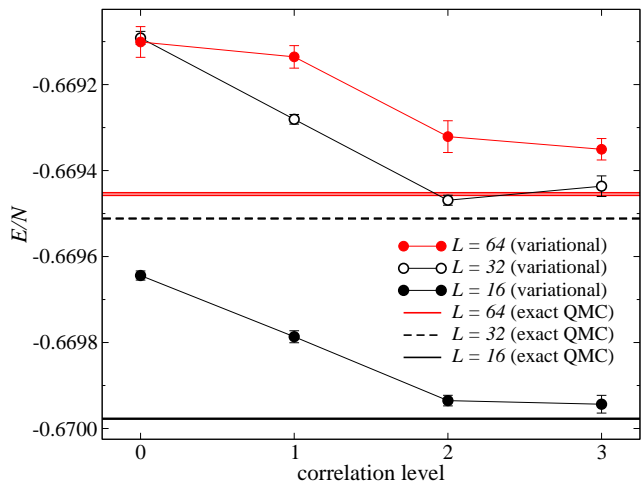}
\caption{(Color online) Energy of the 2D Heisenberg model with variational CAP states at three different levels of bond correlations, according
to the definition of the levels in Fig.~\ref{levels}. Level $0$ corresponds to the pure AP state, with no bond correlations included.
The horizontal lines show energies obtained with unbiased QMC calculations (with the width of the lines corresponding approximately
to the statistical errors).} 
\label{ehberg}
\end{figure}

For the 2D Heisenberg model, previous variational AP calculations have shown that the energy error within this class of state is $<0.1\%$ for large
systems, and the spin correlations are reproduces to within $1\%$ or better.\cite{lou07,sandvik10a} Although the system is strongly N\'eel-ordered and only has 
rapidly decaying short-range VBS correlations, including bond correlations with CAP states can still significantly improve the energy further. Fig.~\ref{ehberg}
shows results for $L\times L$ systems with $L=16,32$, and $64$ at different correlation levels. The deviation from unbiased QMC calculations decreases with
increasing correlation level. For $L=16$ with $r_{\rm max}=3$ the relative error is as small as $\approx 4\times 10^{-5}$, while for the larger systems
it is somewhat larger, about $10^{-4}$. 

\begin{figure}
\includegraphics[width=8cm, clip]{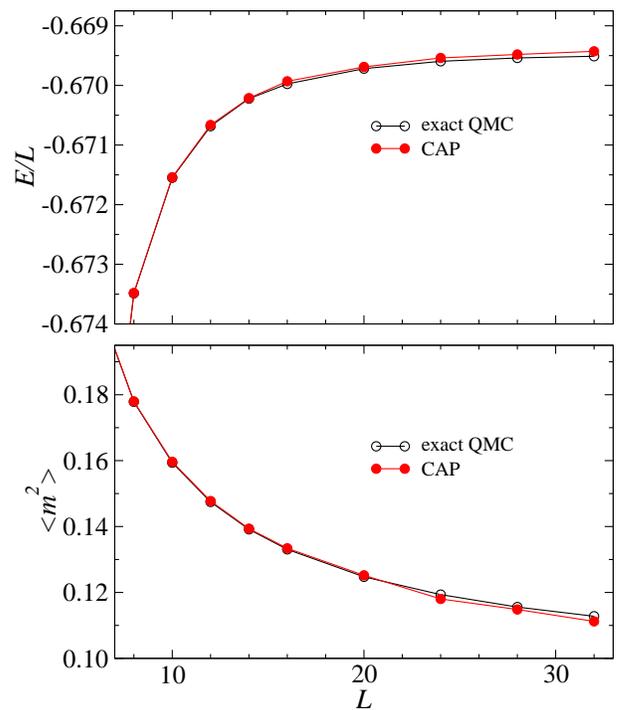}
\caption{(Color online) Energy (top panel) and squared sublattice magnetization (bottom panel) of the 2D Heisenberg model
obtained by unbiased QMC calculations and by optimized CAP state with all bond correlations included.\cite{sandvik10b}} 
\label{es2d}
\end{figure}

Going further and optimizing all correlation weights $C({\bf r}_1,{\bf r}_2)$ with $r \le L/2-1$, one should in principle be able to further improve the energy
and obtain the best possible CAP state (with the kind of correlations included here) when $L\to \infty$. The energy only improves marginally on the $r_{\rm max}=3$
results, however. Fig.~\ref{es2d} shows results versus the system size for the energy as well as the sublattice magnetization. On the scale of the graphs, one 
can barely see any differences between the CAP and unbiased QMC results for $L\le 20$, while for the larger systems there are some visible deviations. Here it 
should again be noted that the results for large systems are likely not completely optimized. As discussed above in Sec.~\ref{sec:optmethod}, 
the energy depends only very weakly on the long-bond statistics, which implies that MC evaluations of the corresponding derivatives are affected by relatively
large fluctuations, leading to slow convergence. The sublattice magnetization is more sensitive to the long bonds, however, and this makes it very difficult
to obtain completely unbiased results for large systems. For example, five independent optimizations for $L=32$ with $r_{\rm max}=3$ all gave the same energy
within statistical errors, but the sublattice magnetization showed significant fluctuations, with the following results: $\langle m^2\rangle = 0.1131,~0.1113, 
0.1132, 0.1129, 0.1094$, with the error bar approximately equal to $2$ in the last digit. Here it can be noted that three of the results agree well, while
two of them are clearly off. One may then conclude that the best optimized results should be around $0.1131$, although to confirm this one should carry
out a much larger number of independent runs. The correct results based on unbiased QMC calculations \cite{sandvik10a} is $\langle m^2\rangle = 0.1128$, 
less that $0.3\%$ below the average of the above $3$ consistent points.

\subsubsection{J-Q model}

As discussed in Sec.~\ref{intro}, the $J$-$Q$ model (\ref{jqham}) exhibits a N\'eel--VBS transition at a critical value of the coupling ratio $J/Q$, with most 
precise estimate so far being $(J/Q)_c=0.0447(2)$.\cite{sandvik10} An interesting question is whether this transition can be described by the CAP states. Here 
we consider the case $J=0$, where the ground state is a columnar VBS. This VBS is very complex, however, since the order parameter is only about $20\%$ of the 
maximum possible value (i.e., for states with length-1 singlets forming columns and no fluctuations around this configuration).\cite{sandvik12} The fluctuations
are significant and have U($1$) character up to a very large length scale (larger than what can currently be studied). As it turns out, the fully optimized
CAP state (using $r_{\rm max}=L/2-1$) for this $J=0$ system does not reproduce the VBS order. Instead, as we will see below, the system is still on the N\'eel 
side of the quantum phase transition. 

\begin{figure}
\includegraphics[width=8.4cm, clip]{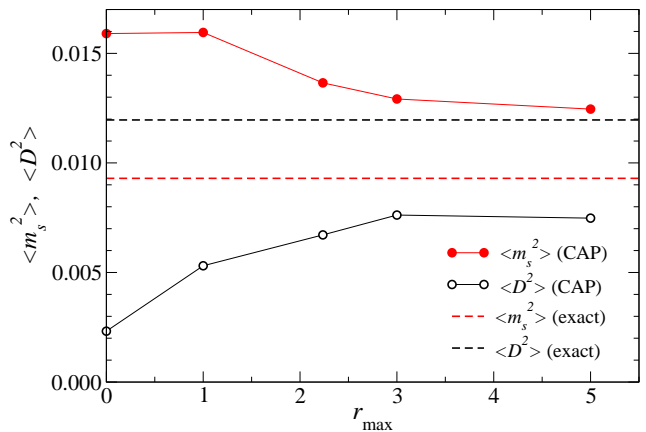}
\caption{(Color online) Squared antiferromagnetic and VBS order parameters versus the correlation level in CAP states
optimized for the $Q$ model on a $32\times 32$ lattice.}
\label{q32}
\end{figure}

First, let us again investigate the impact of including longer bonds in the correlation factor in Eq.~(\ref{psicap}). The energy is improved very dramatically 
with increasing $r_{\rm max}$. For example, for $L=32$ the best optimized AP state has an energy $E_0/N=-0.8013$. The CAP states have $E/N=-0.8215$, $-0.8229$,
and $-0.8232$, at correlation levels $1,2,3$ in the scheme of Fig.~\ref{levels}. Going to $r_{\rm max}=5$ there is only a marginal energy improvement to 
$-0.8233$, which differs by about $0.1\%$ from an unbiased QMC result; $E/N=-0.8240$. 

Although an energy deviation of $0.1\%$ would normally be considered excellent in variational calculations, the order parameters are still not well reproduced.
Fig.~\ref{q32} shows the dependence of both the N\'eel and VBS order parameters on $r_{\rm max}$. With increasing $r_{\rm max}$, the sublattice magnetization is
reduced and the VBS order parameter increases, as would be expected with CAP states in a VBS state. However, the VBS order parameter is still more than $30\%$
too small at $r_{\rm max}=5$, and it appears to be essentially converged at that point. Accordingly, the N\'eel order parameter is instead too large.

\begin{figure}
\includegraphics[width=8.4cm, clip]{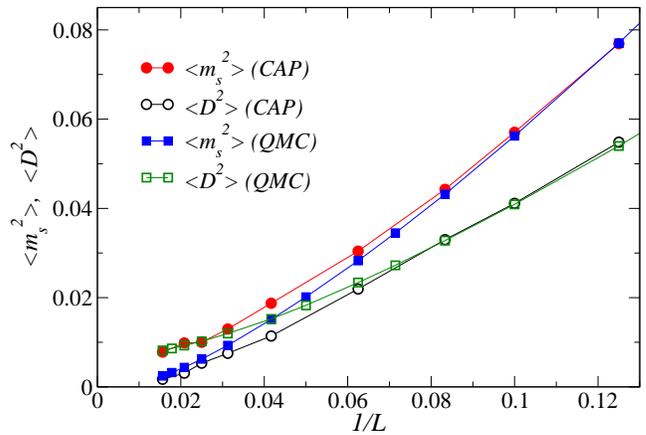}
\caption{(Color online) Squared antiferromagnetic and VBS order parameters versus the inverse system length for the
2D $Q$ model within the CAP states with $r_{\rm max}=3$.}
\label{qorder}
\end{figure}

Fig.~\ref{qorder} shows both order parameters calculated with $r_{\rm max}=3$ as a function of the inverse system size, along with unbiased QMC results. Here one
can observe that the agreement between the two calculations is very good for small systems, but the agreement becomes worse for increasing $L$. Asymptotically, 
the variational calculations tend toward a weakly N\'eel ordered state, with no VBS long-range order, as opposed to the actual VBS ground state. This calculation 
serves to illustrate the insensitivity of the energy to long-distance correlations and the related difficulty in using the energy as a reliable measure of the 
quality of a state obtained by variational means (see Ref.~\onlinecite{liu10} for a different example of this issue).

While this result could be seen as a failure of the variational CAP states, it should be noted again that the actual N\'eel--VBS transition takes place at a very
small coupling ratio, $(J/Q)_c \approx 0.045$, and one cannot expect a variational calculation to reproduce the critical point exactly. For the CAPs considered here 
we have confirmed that the transition is pushed to a small negative $J/Q$, but we leave more detailed studies of the transition (which requires very well optimized 
states) for a future study.

\section{Summary and discussion}
\label{sec:summary}

In conclusion, we have discussed VBS states and associated quantum phase transitions in 1D and 2D wave functions in the VB basis. VBS states appear naturally 
within the standard 1D AP states, and we have here characterized the continuous N\'eel-VBS transition in such a class of states with amplitudes decaying as a power-law 
of the bond length. To stabilize a 2D VBS requires explicit bond correlations beyond the AP states. We have introduced the CAP states, where correlations are enforced 
through factors corresponding to bond pairs, as in the wave function (\ref{psicap}). We have shown how tuning of parameters in 2D CAP states can lead to transitions 
from the standard N\'eel antiferromagnet to a VBS, in some cases with an intervening spin liquid phase. With the parametrization considered here, the direct N\'eel--VBS 
transition is first-order, while the N\'eel--liquid and liquid--VBS transitions are continuous. 

The 2D  N\'eel--liquid transition is of the same kind as in the pure AP states, although the short-distance VBS fluctuations are enhanced. Interestingly, the liquid--VBS 
transition is not associated with an emergent U($1$) symmetry, although the columnar VBS (which is the VBS variant stabilized with the CAPs studied here) in principle 
supports this phenomenon \cite{senthil04a} and has been observed in QMC studies of $J$-$Q$ models at the N\'eel--VBS transition.\cite{sandvik07,jiang08,sandvik12} 
Thus, in the states we have studied here the VBS should be induced by a relevant operator, instead of the dangerously irrelevant operator associated with the emergent 
U($1$) symmetry in the ``deconfined'' criticality scenario.

It remains an interesting challenge to find a parametrization of the CAP states such that a DQC point is obtained. In the study with short-bond correlations in 
Sec.~\ref{sec:2d} we kept the single-bond amplitudes fixed with the form $h(r) = 1/r^3$ and varied only a correlation parameter $p$. In principle we could also
use $h(r) = 1/r^\kappa$ and also vary $\kappa$. It would be interesting to study the full phase diagram in the plane $(p,\kappa)$ for the two parametrizations
of the bond correlations, and also with different choises of the correlation factors.

We note a recent study of AP states, with a certain parametrization of the amplitudes, to describe the N\'eel to quantum-paramagnetic transition in the bilayer 
Heisenberg model.\cite{liao11} The correct 3D classical Heisenberg exponents were obtained for this transition, i.e., the AP states contain effectively long-range 
interactions that allow $(2+1)$-dimensional criticality to be reproduced within a 2D configuration space. In principle it seems that this should be possible 
to achieve also for the DQC transition, and if such a program to construct a simple CAP wave function is successful, it would likely lead to further useful 
insights into the mechanism of spinon deconfinement and emergence of the effective U($1$) gauge field.

It is possible that CAP states with all parameters adjusted to minimize the energy of a model such as the $J$-$Q$ model could lead to a DQC point. Here
we carried out such variational QMC calculations for the standard $J$-$Q$ model with four-spin interactions. This model has a critical point very close to $J/Q=0$,
however, and in the variational calculation the $J=0$ system is still inside the N\'eel phase. In principle one can still study the phase transition by going to 
negative $J$, but in that case Marshall's sign rule cannot be proven rigorously (although most likely it should still hold when $J/Q$ is small and negative). We
will investigate this case in future studies, and also consider the model with six-spin interactions,\cite{lou09}  where the transition point is at much larger $J/Q$ 
and should remain well within the range of positive $J/Q$ values within optimized CAP states. If indeed the correct type of criticality can be achieved, then by
examining the bond amplitudes and correlation factors in the optimized state it may also be possible to construct a class of CAP states with a single tunable 
parameter (instead of all the parameters changing as a function of $J/Q$ in the variationally optimized states) to drive this type of criticality---which the 
parametrization used in the present paper was not capable of.

Beyond the case (I) and (II) parametrizations of the CAP states that we considered here, we have also explored other cases. In particular, with
$c_1=c_3=p$ and $c_2=c_4=1/p$, in the notation of Fig.~\ref{w1234}, we have found a plaquette VBS, in contrast to the columnar VBS states obtaining
in cases (I) and (II). In future studies it will also be interesting to study the nature of the transition from a N\'eel antiferromagnet to a VBS
in this case.

It is in principle possible to further improve on the CAP states, by including more correlations. We here considered pairs of bonds connected to a nearest-neighbor
link only. We have verified that this gives a better variational energy (for the Heisenberg and $J$-$Q$ models) than next-nearest-neighbor links, and therefore, most
likely, the nearest-neighbor links are optimal for introducing correlations in this kind of CAP states. One could also combine several types of correlation factors, 
and include also factors for correlations between more than two bonds. In practice this may not be worth the effort, however, as the main utility of CAP states 
should be (i) to have a simple class of states to capture the N\'eel--VBS transition and (ii) to use them as ``trial states'' for projector QMC calculations in the 
VB basis.\cite{liang90,santoro99,sandvik05,sandvik10a} While the variational states can be improved, in practice it is better to project out the ground state exactly 
using QMC if completely unbiased results are needed, and too many parameters in a CAP state defeats the purpose of (i).

A very interesting question is whether the phase transitions we have discussed here can be realized in ground states of reasonable Hamiltonians,
with only local interactions. In particular, the N\'eel-liquid-VBS series of phases would be of interest in this regard. We already know from the
worh of Cano and Fendley on the short-bond RVB that there is a local parent Hamiltonian for that state. It is then also appears very plausible that 
some local pertutbations of this Hamiltonian will effect the stiffness constant characterizing the RVB \cite{tang11b,damle12} and governing its critical 
VBS correlations. Thus, a class of local Hamiltonians should be able to capture the whole spin liquid phase in our case (II). Then, it also seems plausible 
that other local perturbations can drive the system into a N\'eel or VBS states, e.g., the $Q$ term of the $J$-$Q$ model should do this. Since the Cano-Fendley
Hamiltonian has a sign problem in QMC calculations, some other methods would be needed to study phase transitions in perturbations of it.

\begin{acknowledgments}
We would like to thank Ribhu Kaul for useful comments and discussion.
This work was supported by the NSC under Grant No.~98-2112-M-004-002-MY3 (YCL) and by the NSF under Grant No.~DMR-1104708 (AWS). YCL would like to thank the 
Condensed Matter Theory Visitors Program at Boston University for support and AWS gratefully acknowledges support from the NCTS in Taipei
for visits to National Chengchi University.
\end{acknowledgments}

\null


\vskip-6mm
\begin{thebibliography}{00}

\bibitem{pauling33}
L. Pauling, J. Chem. Phys. {\bf 1}, 280 (1933).

\bibitem{hulthen38}
L. Hulth\'en, Ark. Mat. Astron. Fys. {\bf 26A}, 1 (1938).

\bibitem{sutherland88}
B. Sutherland, Phys. Rev. B {\bf 37}, 3786 (1988); Phys. Rev. B 38, 6855 (1988)

\bibitem{liang88}
S. Liang, B. Doucot, and P. W. Anderson, Phys. Rev. Lett. {\bf 61}, 365 (1988).

\bibitem{beach06}
K. S. D. Beach and A. W. Sandvik, Nucl. Phys. B \textbf{750}, 142 (2006).

\bibitem{wildeboer11}
J. Wildeboer and A. Seidel, Phys. Rev. B {\bf 83}, 184430 (2011). 

\bibitem{fazekas74}
P. Fazekas and P. W. Anderson, Philos. Mag. {\bf 30}, 423 (1974).

\bibitem{shastry81}
B. S. Shastry and B. Sutherland, Phys. Rev. Lett. {\bf 47}, 964 (1981). 

\bibitem{anderson87}
P. W. Anderson, Science {\bf 235} 1196 (1987).

\bibitem{read89a}
N. Read and B. Chakraborty, Phys. Rev. B {\bf 40}, 7133 (1989).

\bibitem{bonesteel89}
N. E. Bonesteel, Phys. Rev. B \textbf{40}, 8954 (1989).

\bibitem{beach09}
K. S. D. Beach, Phys. Rev. B {\bf 79}, 224431 (2009).

\bibitem{wang10}
L. Wang and A. W. Sandvik, Phys. Rev. B {\bf 81}, 054417 (2010).

\bibitem{banerjee10}
A. Banerjee and K. Damle,  J. Stat. Mech. {\bf 2010}, P08017.

\bibitem{banerjee11}
A. Banerjee, K. Damle, and F. Alet, Phys. Rev. B {\bf 83}, 235111 (2011). 

\bibitem{tran11}
H. Tran and N. E. Bonesteel, Phys. Rev. B {\bf 84}, 144420 (2011).

\bibitem{tang11a}
Y. Tang and Anders W. Sandvik,
Phys. Rev. Lett. {\bf 107}, 157201 (2011).

\bibitem{marshall55}
W. Marshall, Proc. Roy. Soc. A {\bf 232}, 48 (1955).

\bibitem{sandvik10a}
A. W. Sandvik and H. G. Evertz, Phys. Rev. B. {\bf 82}, 024407 (2010).

\bibitem{lou07}
J. Lou and A. W. Sandvik, Phys. Rev. B {\bf 76}, 104432 (2007).

\bibitem{tang11b}
Y. Tang, A. W. Sandvik, and C. L. Henley, Phys. Rev. B {\bf 84}, 174427 (2011).

\bibitem{albuquerque10}
A. F. Albuquerque and F. Alet, Phys. Rev. B {\bf 82}, 180408 (2010).

\bibitem{ju12}
H. Ju, A. B. Kallin, P. Fendley, M. B. Hastings, and R. G. Melko, Phys. Rev. B {\bf 85}, 165121 (2012).

\bibitem{cano10}
J. Cano and P. Fendley, Phys. Rev. Lett. {\bf 105}, 067205 (2010).

\bibitem{havilio99}
M. Havilio and A. Auerbach, Phys. Rev. Lett. {\bf 83}, 4848 (1999).

\bibitem{senthil04a}
T. Senthil, A. Vishwanath, L. Balents, S. Sachdev, and M. P. A. Fisher, Science \textbf{303}, 1490
(2004). 

\bibitem{senthil04b}
T. Senthil, L. Balents, S. Sachdev, A. Vishwanath, and M. P. A. Fisher, Phys. Rev. B {\bf 70}, 144407 (2004).

\bibitem{chakravarty89}
S. Chakravarty, B. I. Halperin, and D. R. Nelson, Phys. Rev. B {\bf 39}, 2344 (1989).

\bibitem{read89b}
N. Read  and S. Sachdev, Phys. Rev. Lett. {\bf 62}, 1694 (1989).

\bibitem{murthy90}
G. Murthy and S. Sachdev, Nucl. Phys. B {\bf 344} (1990).

\bibitem{chubukov94}
A. V. Chubukov, S. Sachdev, and J. Ye, Phys. Rev. B {\bf 49}, 11919 (1994).

\bibitem{motrunich04}
O. I. Motrunich and A. Vishwanath, Phys. Rev. B {\bf 70}, 075104 (2004).

\bibitem{sandvik07}
A. W. Sandvik, Phys. Rev. Lett. {\bf 98}, 227202 (2007).

\bibitem{melko08}
R. G. Melko and R. K. Kaul, Phys. Rev. Lett. {\bf 100}, 017203 (2008);
R. K. Kaul and R. G. Melko, Phys. Rev. B {\bf 78}, 014417 (2008).

\bibitem{jiang08}
F. J. Jiang, M. Nyfeler, S. Chandrasekharan, and U. J.  Wiese, J. Stat. Mech. {\bf 2008}, P02009.

\bibitem{lou09}
J. Lou, A. W. Sandvik, and N. Kawashima, Phys. Rev. B {\bf 80}, 180414(R) (2009).

\bibitem{sandvik11}
A. W. Sandvik, V. N. Kotov, and O. P. Sushkov, Phys. Rev. Lett. {\bf 106}, 207203 (2011).

\bibitem{sandvik10}
A. W. Sandvik, Phys. Rev. Lett. {\bf 104}, 177201 (2010)

\bibitem{kaul12}
R. K. Kaul and A. W. Sandvik, Phys. Rev. Lett. {\bf 108}, 137201 (2012).

\bibitem{sandvik06}
A. W. Sandvik and R. Moessner, Phys. Rev. B {\bf 73} 144504 (2006).

\bibitem{mermin66}
N. D. Mermin and H. Wagner, Phys. Rev. Lett. {\bf 17}, 1133 (1966). 

\bibitem{binder81}
K. Binder, Z. Phys. B: Condens. Matter {\bf 43}, 119 (1981).

\bibitem{rokhsar88}
D. S. Rokhsar and S. A. Kivelson, Phys. Rev. Lett. {\bf 61}, 2376 (1988).

\bibitem{affleck85}
I. Affleck, Phys. Rev. Lett. {\bf 55}, 1355 (1985).

\bibitem{korepin04}
V. E. Korepin, Phys. Rev. Lett. {\bf 92}, 096402 (2004).

\bibitem{calabrese04}
P. Calabrese and J. Cardy, J. Stat. Mech. {\bf 2004}, P06002.

\bibitem{kallin09}
A. B. Kallin, I. Gonz\'alez, M. B. Hastings, and R. G. Melko, Phys. Rev. Lett. {\bf 103}, 117203 (2009).

\bibitem{vollmayr93}
K. Vollmayr, J. D. Reger, M. Scheucher, and K. Binder, Z. Phys. B {\bf 91}, 113 (1993).

\bibitem{continentino04}
M. A. Continentino and A. S. Ferreira, Physica A {\bf 339}, 461 (2004)

\bibitem{sen10}
A. Sen and A. W. Sandvik, Phys. Rev. B {\bf 82}, 174428 (2010).

\bibitem{banerjee11b}
A. Banerjee, K. Damle, and A. Paramekanti, Phys. Rev. B {\bf 83}, 134419 (2011).

\bibitem{sandvik12}
A. W. Sandvik, Phys. Rev. B {\bf 85}, 134407 (2012)

\bibitem{cepas05}
O. C\'epas, A. P. Young, and B. S. Shastry, Phys. Rev. B {\bf 72}, 184408 (2005).

\bibitem{jin12}
S. Jin, A. Sen, and A. W. Sandvik, Phys. Rev. Lett. {\bf 108}, 045702 (2012).

\bibitem{damle12}
K. Damle, D. Dhar, and K. Ramola, Phys. Rev. Lett. {\bf 108}, 247216 (2012). 

\bibitem{sandvik10b}
A. W. Sandvik, AIP Conf. Proc. {\bf 1297}, 135 (2010); arXiv:1101.3281.

\bibitem{logpapers}
I. Affleck, D. Gepner, H. J. Schulz, and T. Ziman, J. Math. Phys. A: Math. Gen. {\bf 22}, 511 (1989);
R. R. P. Singh, M. E. Fisher, and R. Shankar, Phys. Rev. B {\bf 39}, 2562 (1989);
T. Giamarchi and H. J. Schulz, Phys. Rev. B {\bf 39}, 4620 (1989).

\bibitem{liu10}
C. Liu, L. Wang, A. W. Sandvik, Y.-C. Su, and Y.-J. Kao, Phys. Rev. B {\bf 82}, 060410 (2010). 

\bibitem{liao11}
H. Liao and T. Li, J. Phys. Cond. Matt. {\bf 23}, 475602 (2011).

\bibitem{liang90}
S. Liang, Phys. Rev. B {\bf 42}, 6555 (1990).

\bibitem{santoro99}
G. Santoro, S. Sorella, L. Guidoni, A. Parola, and E. Tosatti, 
Phys. Rev. Lett. {\bf 83}, 3065 (1999).

\bibitem{sandvik05}
A. W. Sandvik, Phys. Rev. Lett. {\bf 95}, 207203 (2005)

\end{thebibliography}
\end{document}